\documentclass[%
reprint, 
superscriptaddress,
nofootinbib,
amsmath,amssymb,
aps, prd,
]{revtex4-2}

\usepackage{graphicx}
\usepackage{dcolumn}
\usepackage{bm}
\usepackage[unicode,pdfauthor={S. Shashank et al.}, pdftitle={X-ray reflection from GRMHD}]{hyperref}
\hypersetup{colorlinks,urlcolor=blue,citecolor=blue,linkcolor=blue}

\usepackage{gensymb}
\usepackage{orcidlink}
\usepackage[utf8]{inputenc}
\usepackage[T1]{fontenc}

\newcommand{\be}{\begin{eqnarray}}
\newcommand{\ee}{\end{eqnarray}}
\newcommand{\ba}{\begin{aligned}}
\newcommand{\ea}{\end{aligned}}

\newcommand{\pone}{\textsc{Paper I}}

\begin{document}

\title{Measuring black hole spins with x-ray reflection spectroscopy:\\A GRMHD outlook}

\author{Swarnim~Shashank\,\orcidlink{0000-0003-3402-7212}}
\email{swarnim@fudan.edu.cn}
\email{swarrnim@gmail.com}
\affiliation{Center for Astronomy and Astrophysics, Center for Field Theory and Particle Physics, and Department of Physics,
Fudan University, Shanghai 200438, China}

\author{Askar~B.~Abdikamalov\,\orcidlink{0000-0002-7671-6457}}%
\affiliation{School of Humanities and Natural Sciences, New Uzbekistan University, Tashkent 100001, Uzbekistan}
\affiliation{Ulugh Beg Astronomical Institute, Tashkent 100052, Uzbekistan}

\author{Honghui~Liu\,\orcidlink{0000-0003-2845-1009}}
\affiliation{Institut f\"ur Astronomie und Astrophysik, Eberhard-Karls Universit\"at T\"ubingen, D-72076 T\"ubingen, Germany}

\author{Abdurakhmon~Nosirov\,\orcidlink{0009-0009-0287-2535}}
\affiliation{Center for Astronomy and Astrophysics, Center for Field Theory and Particle Physics, and Department of Physics,
Fudan University, Shanghai 200438, China}

\author{Cosimo~Bambi\,\orcidlink{0000-0002-3180-9502}}
\email[Corresponding author: ]{bambi@fudan.edu.cn}
\affiliation{Center for Astronomy and Astrophysics, Center for Field Theory and Particle Physics, and Department of Physics,
Fudan University, Shanghai 200438, China}
\affiliation{School of Humanities and Natural Sciences, New Uzbekistan University, Tashkent 100001, Uzbekistan}

\author{Indu~K.~Dihingia\,\orcidlink{0000-0002-4064-0446}}
\affiliation{Tsung-Dao Lee Institute, Shanghai Jiao-Tong University, Shanghai 201210, China}

\author{Yosuke~Mizuno\,\orcidlink{0000-0002-8131-6730}}
\affiliation{Tsung-Dao Lee Institute, Shanghai Jiao-Tong University, Shanghai 201210, China}
\affiliation{School of Physics and Astronomy, Shanghai Jiao-Tong University, Shanghai 200240, China}
\affiliation{Institut f\"ur Theoretische Physik, Goethe Universit\"at, D-60438 Frankfurt am Main, Germany}
\affiliation{Key Laboratory for Particle Physics, Astrophysics and Cosmology, Shanghai Key Laboratory for Particle Physics and Cosmology, Shanghai Jiao-Tong University, Shanghai 200240, China}

\date{\today}

\begin{abstract}
X-ray reflection spectroscopy has evolved as one of the leading methods to measure black hole spins. However, the question is whether its measurements are subjected to systematic biases, especially considering the possible discrepancy between the spin measurements inferred with this technique and those from gravitational wave observations. In this work, we use general relativistic magnetohydrodynamic (GRMHD) simulations of thin accretion disks around spinning black holes for modeling the accretion process, and then we simulate NuSTAR observations to test the capability of modern reflection models in recovering the input spins. For the first time, we model the electron density and ionization profiles from GRMHD-simulated disks. Our study reveals that current reflection models work well only for fast-rotating black holes. We model the corona as the base of the jet and we find that reflection models with lamppost emissivity profiles fail to recover the correct black hole spins. Reflection models with broken power-law emissivity profiles perform better. As we increase the complexity of the simulated models, it is more difficult to recover the correct input spins, pointing toward the need to update our current reflection models with more advanced accretion disks and coronal geometries.
\end{abstract}

\keywords{Accretion disk \& black-hole plasma, Astronomical black holes, Astrophysical \& cosmological simulations, x-ray astronomy}

\maketitle

\section{Introduction}\label{sec:intro}

Einstein's theory of general relativity (GR) predicts the existence of black holes. Astrophysical black holes are believed to be approximated well by the Kerr solution~\cite{Kerr:1963ud}; i.e., they have only two properties: the mass and the spin angular momentum (hereafter spin). With the advent of advanced observatories, we have observed black holes in the whole electromagnetic band~\cite{2022iSci...25j3544L, book17, Bambi:2020jpe, EventHorizonTelescope:2021dvx,McClintock:2013vwa} as well as through gravitational waves~\cite{LIGOScientific:2016aoc}.
Accretion onto black holes is an extremely efficient mechanism for converting rest-mass energy into radiation, powering the most luminous astrophysical sources and offering a unique test bed for exploring fluids, magnetic fields, and radiative processes under extreme gravity.
Measuring black hole spins has significant astrophysical importance; see, for example, Ref.~\cite{Reynolds:2020jwt} for a review. Spins dictate the efficiency of how the accreting matter converts its rest-mass energy into radiation~\cite{Bardeen:1972fi}. Power outputs of relativistic jets, which are launched from these systems, may also depend on the spins~\cite{Blandford:1977ds}. In supermassive black holes, spins carry the history of how black holes formed and grew~\cite{Chen:2025trg}. Black hole spins may also be responsible for precessions of disks and jets~\cite{Bardeen:1975zz}.

Theoretical formulations of accretion disks around black holes were laid out five decades ago. Shakura and Sunyaev formulated the radial and vertical structure of disks with angular momentum transport, which are now colloquially known as the standard disk model~\cite{Shakura:1972te,Lynden-Bell:1974vrx,Pringle:1981ds}. Novikov and Thorne developed the relativistic version of the Shakura-Sunyaev model~\cite{1973blho.conf..343N,Page:1974he} (hereafter Novikov-Thorne disk). The Novikov-Thorne model describes an accretion disk in the equatorial plane of a black hole rotating with Keplerian velocity and with an inner edge lying on the innermost stable circular orbit (ISCO)\footnote{We note that Novikov-Thorne disks are thermally and viscously unstable in the range 0.01-1~$L_{\rm Edd}$, where $L_{\rm Edd}$ is the Eddington luminosity of the source~\cite{Lightman:1974sm}. These disks can be stabilized by large-scale magnetic fields~\cite{Begelman:2006vj,Lancova:2019ldt}, which, however, alter the disk structure.}. These analytical models rely on an artificial viscosity treatment called the $\alpha$ prescription for the angular momentum transport (and for this reason they are often also called $\alpha$ disks). It has been shown that this angular momentum transport is carried out from the magnetorotational instability resulting from a magnetohydrodynamic treatment of the disk~\cite{Balbus:1991ay,Balbus:1998ja,Balbus:1999fk}. In this regard, the development of general relativistic magnetohydrodynamics (GRMHD) has proven to reveal unprecedented details about the accretion process (see Ref.~\cite{Bambi:2025btf} and the chapters therein for a recent review). 

\begin{figure}
    \centering
    \includegraphics[width=0.95\linewidth]{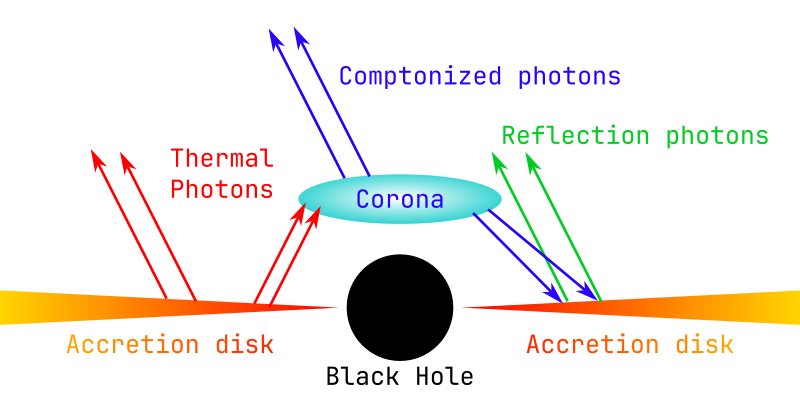}
    \caption{A sketch of the disk-corona model.}
    \label{fig:disk-corona}
\end{figure}

In Fig.~\ref{fig:disk-corona}, we show the disk-corona model, which consists of a central black hole, a cold accretion disk, and a hot corona.
In the disk-corona model, the spectra of accreting black holes come in three flavors. $i)$~A thermal continuum from a geometrically thin and optically thick disk (red arrows in Fig.~\ref{fig:disk-corona})\footnote{We note that the disk-corona model does not require that the inner edge of the disk is at the ISCO. The disk may indeed be {\it truncated} at a radius larger than the ISCO. }: since the disk temperature roughly scales as $M^{-0.25}_{\rm BH}$, where $M_{\rm BH}$ is the black hole mass, the continuum is peaked in the soft x-ray band in the case of stellar-mass black holes in x-ray binaries and in the UV band in the case of supermassive black holes in active galactic nuclei. $ii)$~A Comptonized spectrum produced by the inverse Compton scattering of thermal photons of the disk off free electrons in the hot corona (blue arrows in Fig.~\ref{fig:disk-corona}). $iii)$~A reflection spectrum, which is the result of the reprocessing of the Comptonized photons by the disk (green arrows in Fig.~\ref{fig:disk-corona}). The reflection spectra harbor fluorescent emission lines, which are broadened due to relativistic effects~\cite{Fabian:1989ej,Tanaka:1995en,Nandra:2007rp,Miller:2009cw}. The main features in a reflection spectrum are often the Fe K$\alpha$ complex, which is located at 6.4~keV for neutral or weakly ionized iron atoms and at 6.97~keV for H-like iron ions, and a Compton hump peaking around 20-40~keV~\cite{Bambi:2020jpe}.

x-ray reflection spectroscopy refers to the analysis of these relativistically broadened reflection features in the x-ray spectra of accreting black hole. X-ray reflection spectroscopy is currently the only technique to be able to measure the spins for both stellar-mass and supermassive black holes~\cite{Reynolds:2013qqa,Bambi:2020jpe}. For stellar-mass black holes, the other techniques to measure spins are the continuum-fitting method~\cite{McClintock:2013vwa} and gravitational waves \cite{GWTC1,GWTC2,GWTC3,GWTC4}. For supermassive black holes, x-ray reflection is currently the only full-fledged technique to measure black hole spins. There is the possibility of measurement with Event Horizon Telescope (EHT) imaging~\cite{Moriyama:2019mhz,EventHorizonTelescope:2019pgp,EventHorizonTelescope:2022urf} and with future photon ring measurements~\cite{Broderick:2021ohx}, and also with gravitational wave observations with future space-based interferometers~\cite{LISA:2022yao,Li:2024rnk,Wang:2019ryf,Ruan:2018tsw}. 
Since x-ray reflection spectroscopy finds that most stellar-mass black holes in x-ray binaries are fast-rotating while gravitational waves find that most stellar-mass black holes in binary black holes are slow-rotating, it is now under debate if this is because black hole x-ray binaries and binary black holes are two different object classes, whether there are observational biases, or at least one of the two methods does not provide accurate black hole spin measurements~\cite{Reynolds:2020jwt,Zdziarski:2025ozs}.

A crucial ingredient in the spin measurements with x-ray reflection spectroscopy is that the inner edge of the disk is at the ISCO. While there is a body of evidence that this is indeed the case in x-ray binaries when the spectrum of a source is dominated by the thermal component of the disk (soft state)~\cite{Steiner:2010kd,Penna:2010hu,Kulkarni:2011cy}, it is still a controversial issue when the spectrum is dominated by the Comptonized spectrum of the corona and/or the reflection spectrum of the disk (hard state)~\cite{Fabian:2014tda}. When the luminosity of the source is low, the disk is likely truncated in the hard state, while the inner edge of the disk may approach the ISCO radius when the luminosity of the source increases~\cite{Wang-Ji:2017oly,Liu:2023ovm,Fan:2024ixo}. It is sufficient that the disk is truncated at a few gravitational radii that the reflection spectrum is not very sensitive to the black hole spin any longer, precluding an accurate measurement of this parameter. On the other hand, the assumption that the inner edge of the accretion disk is at the ISCO radius is not very important in the case of measurements of very high spins, when the dimensionless spin parameter $a_*$ is estimated to be very close to 1, which is the case of most spin measurements using x-ray reflection spectroscopy~\cite{Bambi:2020jpe,Draghis:2023vzj}: those observations require in any case that the inner edge of the disk is very close to the black hole, which is only possible when the black hole is rotating fast and the inner edge of the disk is close to the ISCO. For this reason, the discrepancy between spin measurements from x-ray reflection spectroscopy and gravitational waves cannot be attributed to the ISCO assumption in reflection modeling.

The recent years have seen a lot of development in x-ray reflection spectroscopy. State-of-the-art models like \texttt{relxill}~\cite{Dauser:2013xv,Garcia:2013lxa}, \texttt{relxill\_nk}~\cite{Bambi:2016sac,Abdikamalov:2019yrr}, \texttt{reltrans}~\cite{Ingram:2019qlb,Mastroserio:2021jyj}, \texttt{kyn}~\cite{Dovciak:2003jym}, and \texttt{reflkerr}~\cite{Niedzwiecki:2018wtc} have been developed and are perpetually ongoing modifications to include new physics. 
There are still many assumptions that these relativistic reflection models rely upon and that need to be understood for precise and accurate measurements of spins when doing data analyses~\cite{Bambi:2020jpe}. These simplifications arise from the following parts: description of the accretion disk, description of the hot corona, approximations in the calculations of rest-frame reflection spectra, and relativistic effects.
There has also been a lot of testing related to different simplifications. For example, some studies have investigated the impact of various assumptions about the accreting material, like disk thickness~\cite{Taylor:2017jep,Abdikamalov:2020oci,Riaz:2019bkv,Shashank:2022xyh}, emissions from plunging regions \cite{Zhou:2019dfw,Cardenas-Avendano:2020xtw}, and super-Eddington regimes \cite{Zhang:2024pzd,Shashank:2024hsi}. Different simplistic corona geometries, like lamppost, ringlike, and disklike coronae, have been tested and implemented~\cite{Dauser:2013xv,Riaz:2020svt,Feng:2025tkh}. Implementation of ionization and density profiles in the local reflection spectrum has also been studied~\cite{Abdikamalov:2021rty,Abdikamalov:2021ues}. Efforts have been made to include relativistic phenomenon, like the returning radiation~\cite{Dauser:2022zwc,Mirzaev:2024fgd,Mirzaev:2024qcu}, emission angle corrections~\cite{Liu:2024xim,Huang:2025fkq}, non-Keplerian rotation~\cite{Tripathi:2020wfi}, and even deviations from GR~\cite{Bambi:2016sac,Abdikamalov:2019yrr,Abdikamalov:2020oci,Tripathi:2020yts}.

GRMHD simulations of accreting black holes have been very popular in the simulations of radiatively inefficient accretion flows, which are an important system for EHT observations~\cite{EventHorizonTelescope:2019pcy,Bambi:2025btf}. In the case of radiatively efficient simulations, such as for the case for systems like black hole x-ray binaries and active galactic nuclei, a radiative cooling mechanism is added to make the disk thin. This has previously been achieved using some \textit{ad hoc} cooling mechanisms~\cite{Noble:2008tm,Penna:2010hu,Avara:2015kna}. Only very recently have more physical cooling mechanisms have been implemented. However, there are still issues related to runaway cooling and thermal instability of the disk that need to be corrected by adding a floor value to the cooling~\cite{Dihingia:2023mng,Motta:2025gza,Jiang:2013aoa}. Recent years have seen many works with modeling of disk emissions using GRMHD simulations of thin accretion disks~\cite{Schnittman:2015nvd,Kinch:2018ceh,Kinch:2021zfx,Liu:2024ykw}, some where the state transitions and disk truncations have been studied~\cite{Liska:2022jdy,Dihingia:2022wav,Liska:2023fbv,Dihingia:2023eav} and few where the iron line emissions have been modeled~\cite{Reynolds:2007rx,Kinch:2016ipi,Nampalliwar:2022smp}. However, our previous work~\cite{Shashank:2022xyh} (hereafter \pone) remains the only one where the full reflection spectrum was modeled and x-ray observations were simulated.

In \pone, we performed a GRMHD simulation of a fast-rotating black hole with a thin accretion disk and then, using ray tracing, we calculated the reflection spectra for high and low inclination angles. The \texttt{xillver} model~\cite{Garcia:2013oma,Garcia:2013lxa} was used to calculate the local reflection spectrum of the disk. After which, we simulated 30~ks observations of bright Galactic sources with NuSTAR~\cite{NuSTAR:2013yza}. The data were fit with the \texttt{relxill}\footnote{\url{https://sternwarte.uni-erlangen.de/~dauser/research/relxill/index.html}} and \texttt{relxill\_nk}\footnote{\url{https://github.com/ABHModels/relxill_nk}} models to check for systematic uncertainties. The main idea for that work was to test the capability of \texttt{relxill} and \texttt{relxill\_nk} to recover the input parameters that were used in the GRMHD simulation and radiation transport. Since simulations model the accretion by solving the GRMHD equations, they can be believed to provide a more advanced description of thin accretion disks than the analytical, Keplerian, infinitesimally thin disk model assumed in current reflection models\footnote{We note that, strictly speaking, current reflection models do not employ Novikov-Thorne disks. They only assume that the disk is Keplerian, infinitesimally thin, and perpendicular to the black hole spin axis. The inner edge of the disk can be a free parameters in these models, even if it is common to impose that it is at the ISCO when we want to measure black hole spins.}.
\pone ~revealed that \texttt{relxill} and \texttt{relxill\_nk} perform well for high inclination disks, while minor discrepancies were found for low inclination disks. 

In this work, we extend the study that was performed in \pone. We start the simulations with a thin disk geometry instead of a torus geometry and choose three spin values (0.5, 0.8, and 0.98). We seed magnetic fields to produce a jet, which is then adopted as the hot corona. Jetlike coronae have been explored in various studies; see, for example, Refs.~\cite{Markoff:2003ay,Markoff:2004wn,Markoff:2005ht,Dauser:2013xv,You:2021khm,Davidson:2025zpj}. We use the time-averaged simulation data to perform ray tracing and we calculate reflection spectra for low ($30 \degree$) and high ($70 \degree$) inclinations of the disk to the observer. We use the response files of NuSTAR to simulate a number of observations. This synthetic observations are fit with the latest refection models in the \texttt{relxill} and \texttt{relxill\_nk} suites.
The goal of our work is to figure out: $(i)$ the capability of the reflection models to recover the input spins, $(ii)$ the capability of the lamppost models to fit data for the emissivity of a jetlike corona, $(iii)$ the capability of the reflection models to fit the data given ionization and density profiles derived from the simulations, and $(iv)$ the role of the reflection from the plunging region, which may spoil the spin measurements.

The article is organized in the following manner. In Sec.~\ref{sec:methods}, we describe our methodology, covering the GRMHD simulations, ray tracing methods, and simulations of x-ray observations. In Sec.~\ref{sec:models}, we describe all the different configurations we used for testing the reflection models.
In Sec.~\ref{sec:results}, we describe our results obtained from the fitting of the synthetic observations from each of our configurations. Finally, in Sec.~\ref{sec:conclusion}, we discuss our results and we summarize our conclusions.

All quantities are defined in the units $c = G_{N} = M_{\rm BH} = 1$ unless stated otherwise. $r_g = G_N M_{\rm BH}/c^2$ is the gravitational radius and $t_g = G_N M_{\rm BH}/c^3$ is the gravitational time for a black hole with mass $M_{\rm BH}$. Greek indices run for both space and time, viz., $(t, r, \theta, \phi)$, and Latin indices run only space, viz., $(r, \theta, \phi)$. The metric signature is taken as $(-, +, +, +)$.

\section{Methods}\label{sec:methods}

\subsection{GRMHD simulation setup}\label{sec:sim_setup}

In order to model a thin disk with a jet, we use the \texttt{Athena++}~\cite{Stone2020} code\footnote{\url{https://github.com/PrincetonUniversity/athena}} to solve the GRMHD equations
\be\ba
\nabla_{\nu} \, (\rho u^{\nu}) &=& 0 , \\
\nabla_{\mu} \, T^{\mu}_{\nu} &=& - S u_{\nu} , \\
\nabla_{\mu} \, ^* \mathbb{F}^{\mu\nu} &=& 0 ,
\ea\ee
where $\rho$ is the fluid's rest-mass density and $u^{\mu}$ is the four-velocity of the gas in the disk. $T^{\mu}_{\nu}$ is the energy-momentum tensor, defined as
\be
T^{\mu\nu} = \rho h u^{\mu}u^{\nu} + (P + b^2/2)g^{\mu\nu} - b^{\mu}b^{\nu} ,
\ee
and
\be
^* \mathbb{F}^{\mu\nu} = b^{\mu}u^{\nu} - b^{\nu}u^{\mu}
\ee
is the dual of the Faraday tensor. $b^{\mu}$ are the Lagrangian magnetic field components and $b^2 = b^{\lambda}b_{\lambda}$. $h$ is the Lagrangian specific enthalpy, defined as
\be h = 1 + \frac{P \tilde{\Gamma}}{\rho(\tilde{\Gamma}-1)} + \frac{b^2}{\rho}.
\ee 
The background spacetime is defined by a Kerr-Schild geometry in polar coordinates. We use an ideal gas equation of state during the evolution; i.e., $P = (\tilde{\Gamma} - 1) e$, where we set $\tilde{\Gamma} = 4/3$ for a radiation dominated disk and to accurately account for the jet and plunging region\footnote{The choice of $\tilde{\Gamma} = 5/3$ or $4/3$ produced similar results in this work.}. $P$ is the pressure and $e$ is the internal energy density of gas.
$S u_{\nu}$ is the source term to add a radiative cooling~\cite{Noble:2008tm} and is defined as
\be\ba
S' &= \frac{P \Omega}{\tilde{\Gamma}-1} \left[ Y - 1 + |Y-1| \right]^{1/2}, \\
S &= S' \exp \left(-\sigma^2/\sigma_{\rm cut}^2 \right),
\ea\ee
where $Y = P/\rho\mathcal{T_*}$, $\sigma = b^2/\rho$, and $\sigma_{\rm cut} = 5$ is chosen to not cool the jet. Additionally, the cooling is also switched off when $h u_t < -1$ to avoid cooling any outflow. The quantity
\be
\mathcal{T_*} = \frac{\pi}{2} \left[ \mathcal{H} r \Omega(r) \right]
\ee
defines the temperature of the disk with scale height $\mathcal{H}$ in Newtonian gravity, where $r$ is the radial coordinate. In our simulations, we set $\mathcal{H} = 0.05$ to keep the disk thin throughout the temporal evolution. $\Omega(r)$ is taken as the relativistic Keplerian velocity for $r>r_{\rm ISCO}$ and as the angular velocity of a freely falling fluid particle for $r<r_{\rm ISCO}$~\cite{Noble:2008tm}.

\begin{figure}
    \centering
    \includegraphics[width=0.95\linewidth]{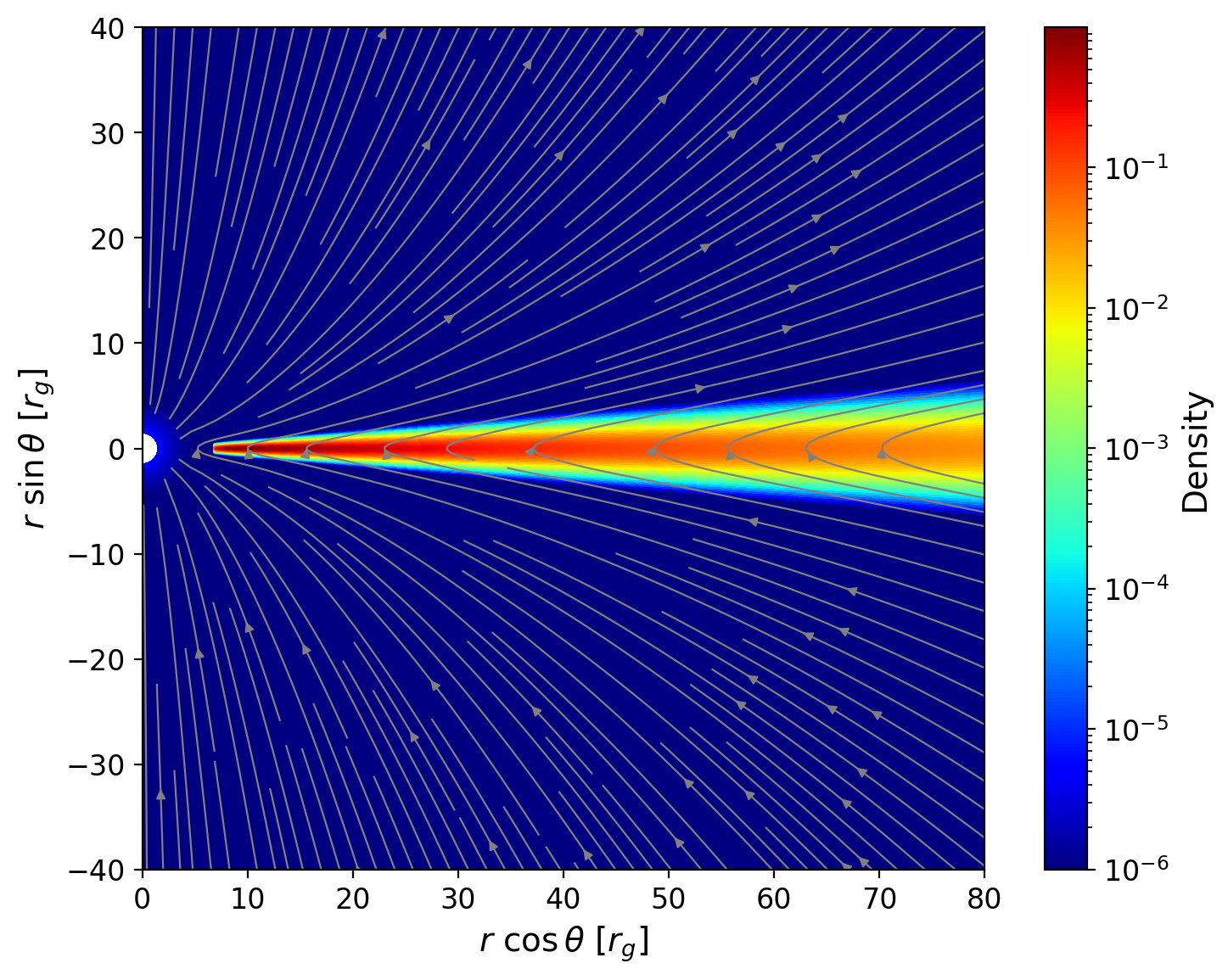}
    \caption{Normalized density profile of the initial configuration of the accretion disk around a black hole of $a_* = 0.5$. The gray lines represent magnetic field lines.}
    \label{fig:initial_disk}
\end{figure}

\begin{figure}
    \centering
    \includegraphics[width=0.95\linewidth]{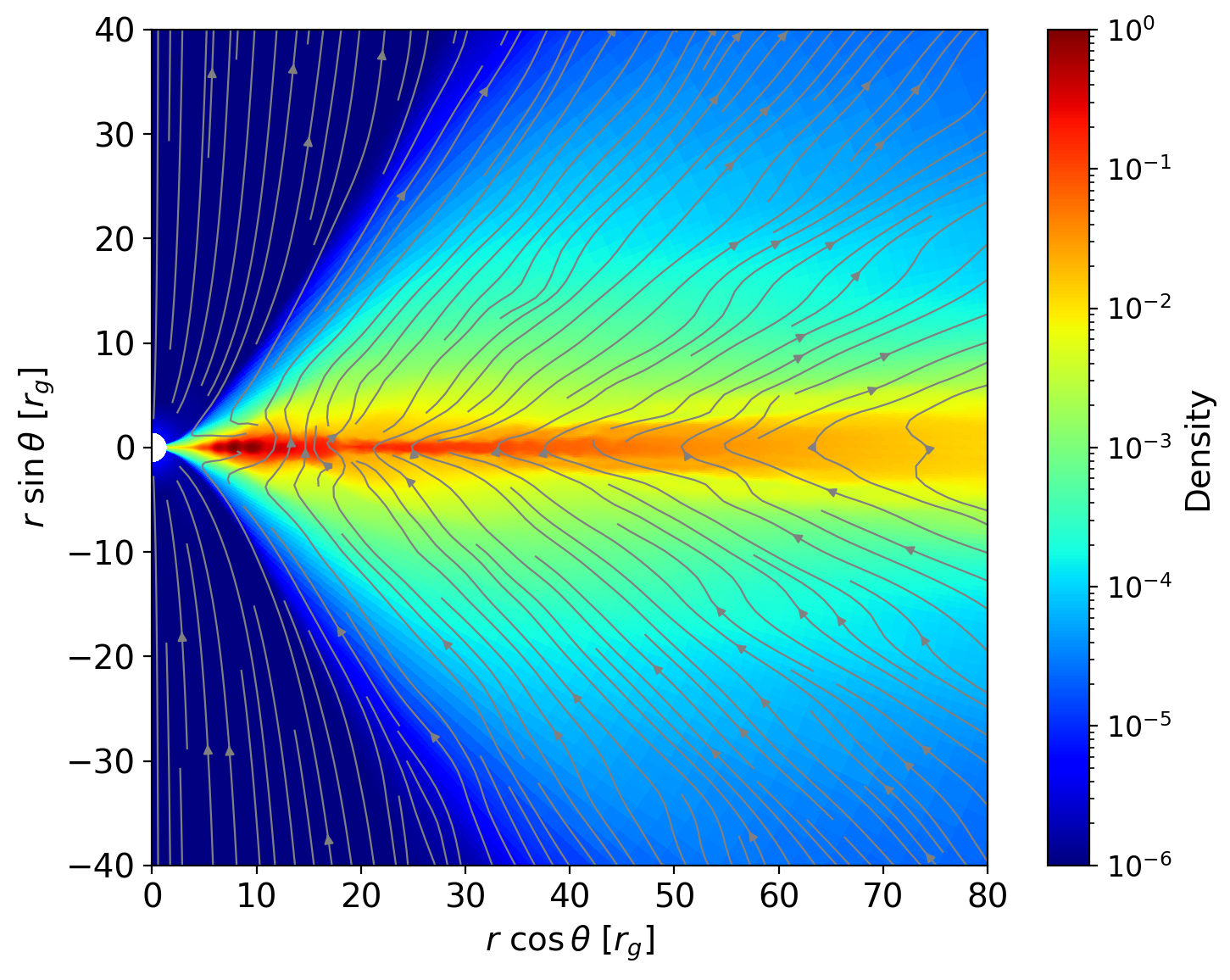}
    \caption{Normalized density profile of the accretion disk around a black hole of $a_* = 0.5$, averaged over $7500~t_g$ -- $15000~t_g$. The gray lines represent magnetic field lines.}
    \label{fig:avg_disk}
\end{figure}

The disk is initialized with a Novikov-Thorne disk at the equatorial plane \cite{Dihingia:2024tqr,Dihingia:2021ncv}, such that
\be
\frac{P_e}{\rho_e} = \Theta_0 \left( \frac{\mathcal{F}}{r} \right)^{1/4} ,
\ee
where the subscript $e$ denotes quantities evaluated in the equatorial plane. Considering a polytropic equation of state $P = K \rho^{\tilde{\Gamma}}$, we define
\be
\rho_e = \left( \frac{\Theta_0}{K} \right)^{\frac{1}{\tilde{\Gamma}-1}} \left( \frac{\mathcal{F}}{r} \right)^{\frac{1}{4(\tilde{\Gamma}-1)}}.
\ee
In our simulations, we take $\Theta_0 = 0.001$, which is a constant related to the disk's initial temperature, and $K = 0.1$, which is the entropy constant~\cite{Dihingia:2024tqr,Dihingia:2021ncv}. In the above equations, $\mathcal{F}$ comes from the time-averaged radiation flux for a Novikov-Thorne disk accreting at a rate $\dot{M}$, given by $F = (\dot{M}/4\pi r)\mathcal{F}$ \cite{Page:1974he}, and can be written in the following form:
\be\ba
\mathcal{F}(x) = \frac{3}{2 x^2 (2a_* + x^3 - 3x)} \left[ x - x_0 - \frac{3}{2} a_* \ln \left( \frac{x}{x_0} \right) \right.\\
- \frac{3 (s_1 - a_*)^2}{s_1 (s_1 - s_2)(s_1 - s_3)} \ln \left( \frac{x - s_1}{x_0 - s_1} \right) \\
- \frac{3 (s_2 - a_*)^2}{s_2 (s_2 - s_1)(s_2 - s_3)} \ln \left( \frac{x - s_2}{x_0 - s_2} \right) \\
\left. - \frac{3 (s_3 - a_*)^2}{s_3 (s_3 - s_1)(s_3 - s_2)} \ln \left( \frac{x - s_3}{x_0 - s_3} \right) \right],
\ea\ee
where
\be\ba
s_1 &= 2 \cos \left( \frac{1}{3} \cos^{-1}a_* - \frac{\pi}{3} \right), \\
s_2 &= 2 \cos \left( \frac{1}{3} \cos^{-1}a_* + \frac{\pi}{3} \right), \\
s_3 &= -2 \cos \left( \frac{1}{3} \cos^{-1}a_* \right),
\ea\ee
$a_*$ is the dimensionless black hole spin parameter, $x^2 = r$, and $x_0^2 = r_{\rm ISCO}$ is the ISCO radius for the given spin.
The initial disk is provided a Gaussian thickness given by
\be
\rho(r, \theta) = \rho_e (r) \exp \left( - \frac{4 \cos^2 \theta}{\mathcal{H}^2} \right),
\ee
where $r$ and $\theta$ are the radial and polar coordinates.
The initial conditions for the four-velocity of the gas can be chosen by assuming each fluid element follows the geodesic equation for a disk in steady state with no radial infall; i.e.,
\be
u^{\mu}u^{\nu} \Gamma_{\mu\nu}^{\lambda} = 0 ,
\ee
and the relation $g_{\mu\nu} u^{\mu}u^{\nu} = -1$.
The final constituent remaining for the initial conditions is the magnetic field. We choose an inclined poloidal field profile, which can be defined by an azimuthal vector potential given by~\cite{Zanni:2007hz,Dihingia:2024tqr}
\be
A_{\phi} \propto (r \sin \theta)^{3/4} \frac{m^{5/4}}{(1 + \tan^{-2} \theta)^{5/8}} .
\ee
Here, $m$ sets the inclination of the field, which we set to $0.1$. We set the plasma parameters $\beta_{\rm min} = 0.5$ and $\sigma_{\rm max} = 50$. The floor values are set to $\rho_{\rm flr} = \rho_{\min}r^{-1.5}$ with $\rho_{\min} = 10^{-5}$ and $P_{\rm flr} = P_{\min}r^{-2.5}$ with $P_{\min} = 1/3 \times 10^{-9}$. The maximum bound of the Lorentz factor is set to $20$. The initial disk is shown in Fig.~\ref{fig:initial_disk}. The simulations are scaleless by default. To derive the physical values, we scale the quantities by considering $M_{\rm BH} = 10~M_{\odot}$ and $\dot{M} = 0.1~\dot{M}_{\rm Edd}$,\footnote{Our definition of Eddington mass accretion rate is $\dot{M}_{\rm Edd} = L_{\rm Edd}/\eta c^2$, where $\eta$ is the radiative efficiency.} where $M_{\odot}$ is the mass of the Sun and $\dot{M}_{\rm Edd}$ is the Eddington accretion rate in CGS units. The simulations are 2.5D with the effective resolution $1024\times1024\times1$ using three levels of static mesh refinement. The radial points are distributed logarithmically, providing a higher resolution in the inner parts of the disk. The range of the radial coordinate is $r \in [0.75~r_{\rm H}, 200~r_g]$ and that of the polar coordinate is $\theta \in (0, \pi)$, where $r_{\rm H}$ is the radial coordinate of the event horizon of the black hole. For the runs, we use the HLLE Riemann solver, third-order strong stability preserving Runge-Kutta for time integration, and third-order piecewise parabolic method for spatial reconstruction. We run the simulations for the spin values $a_* = \{ 0.5, 0.8, 0.98 \}$ until $15000~t_g$. To extract the spectra, we take the time average from $7500~t_g$ to $15000~t_g$. The averaged profile is shown in Fig.~\ref{fig:avg_disk}.

To get the jet profile, we follow the standard method widely used in the literature~\cite{Mizuno:2022vqa}: we require $h u_t < -1$ (unbound fluid) and we take the $\sigma = 1$ contour. We assume that the corona extends from the horizon to the height $\sim 5~r_{\rm g}$, as we want to have a compact corona. Our jet-corona profile is shown in pink in Fig.~\ref{fig:jet-disk}. We note that this is an inflow (particles move toward the black hole) and the outflow starts at a somewhat larger distance from the black hole. However, studies with a two-temperature fluid show that this region is filled with hot electrons~\cite{Dihingia:2023mng} and can thus potentially act as a corona.

\subsection{Ray-tracing setup}\label{sec:rt_setup}

In order to calculate the emissions from the system, it is needed to perform ray tracing and radiative transport. For the ray tracing, we use our publicly available codes \texttt{blacklamp} \cite{abdikamalov_2024_10673757} and \texttt{blackray} \cite{abdikamalov_2024_10673859} modified to account for the GRMHD simulation data. For the radiative transport to calculate the spectra, we use the \texttt{xillver}\footnote{\url{https://sites.srl.caltech.edu/~javier/xillver/index.html}} tables, which are also available in the \texttt{relxill} model. Our ray tracing framework is described in Refs.~\cite{Bambi:2016sac,Abdikamalov:2019yrr,Abdikamalov:2020oci}.

\begin{figure}
    \centering
    \includegraphics[width=0.95\linewidth]{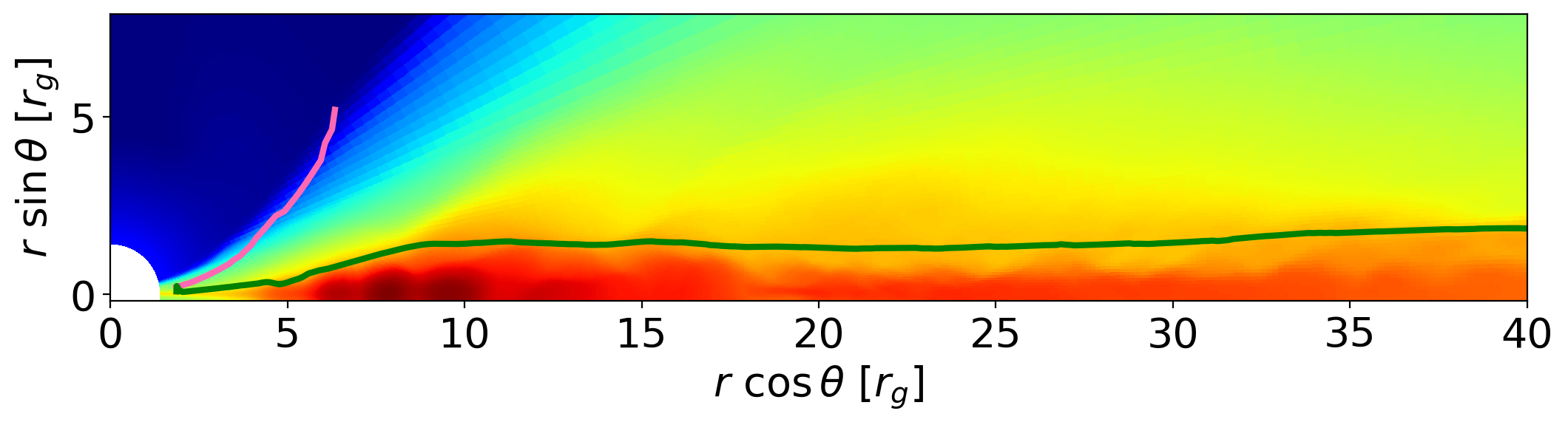}
    \caption{Corona and disk geometry obtained from the simulations and ray tracing for the simulation with black hole spin $a_* = 0.5$. The boundary of the corona is marked by the pink line. The surface of the accretion disk is marked by the green line. Outer edge of the disk for all cases lies $\sim 188~r_g$. The color map is the normalized density as in Fig.~\ref{fig:avg_disk}.}
    \label{fig:jet-disk}
\end{figure}

First, photons are ray traced from the jet corona to the disk (Fig.~\ref{fig:jet-disk} pink line). Indeed we assume that the corona of our system is the jet that we find in our GRMHD simulations (see previous section): this is a finite-size corona and we need to calculate how such a corona illuminates the accretion disk. The optical depth is calculated along the paths of the photons with the following equation
\be
d \tau = \kappa \rho_{\rm cgs} d l_{\rm prop},
\ee
where $d l_{\rm prop} = k_{\mu} u^{\mu} d \lambda$. Here, $k^{\mu}$ is the four-momentum of the photons and $\lambda$ is the affine parameter along the path of the photons. $\rho_{\rm cgs}$ is the density of the gas in CGS units (see previous section) and $\kappa = 0.4 ~\mathrm{cm}^2 \mathrm{g}^{-1}$ is the Thomson scattering opacity.
The integration for the optical depth is stopped when we find an optical depth of $\tau = 1$. This provides the disk surface as seen by the photons where the reflection occurs. The obtained disk is shown in green in Fig.~\ref{fig:jet-disk}. To our knowledge, this is the first time a reflection surface of the disk has been calculated using this method, viz., the reflection surface as seen by the corona obtained using ray tracing. Each coronal emitter sees a disk that is slightly different, so we find an average value of the height of the disk for each radial point.

Second, we calculate how the jet corona\footnote{We choose $\sim 50$ coronal points in a logarithmic distribution i.e. more points near the black hole.} illuminates the accretion disk, viz. the emissivity profile of the disk. For simplicity, we assume the corona to be stationary. We consider 200~radial bins (annuli) along the accretion disk. For each annulus of index $i$, the flux is calculated by
\be
F_{i} \propto N_p\frac{g_c^{\Gamma}}{A_i} ,
\ee
where $N_p$ is the number of photons hitting the annulus $i$, $g_c$ is the average redshift experienced by the photons from the corona to the annulus $i$, $\Gamma$ is the photon index, which we set to 1.7, and $A_i$ is the proper area of the annulus $i$, given by
\be
A_i = 2 \pi \int_{r_i}^{r_i + \Delta r_i} u^t \sqrt{g_{t\phi}^2 - g_{tt} g_{\phi\phi}} ~\sqrt{g_{rr}} dr.
\ee
A full detailed calculation could be found in Ref.~\cite{Bambi:2024afn}. By assuming the total luminosity of the corona to some physical value like $0.01~L_{\rm Edd}$, where $L_{\rm Edd}$ is the Eddington luminosity in CGS units, the obtained flux can be scaled to physical values.

Third, it is required to calculate the redshift and the emission angle at the disk as seen by a distant observer so as to synthesize a full relativistic reflection spectrum. We shoot photons back in time from the observer screen to the accretion disk. When the photons hit the disk, we calculate the redshift and the emission angle on the disk surface calculated earlier. We perform this ray tracing for two inclination angles, low ($i = 30 \degree$) and high ($i = 70 \degree$). The full calculation is given in Refs.~\cite{Bambi:2016sac,Abdikamalov:2019yrr,Abdikamalov:2020oci,Bambi:2024afn}.

Now that we have all ingredients for the reflection spectrum, we can calculate the full reflection spectrum by convolving the local reflection spectrum given by the \texttt{xillverCp} table. Employing the \texttt{xillverCp} table, we assume that the spectrum illuminating every point of the disk is described by the thermally Comptonized continuum model {\tt nthComp}~\cite{Zdziarski:1996wq,Zycki:1999cm}. This is also a simplification because the actual spectrum illuminating the disk is more complicated and different at any radial coordinate. Especially in the hard state, the corona is expected to be extended, so it may not be described well by a model with a single temperature and a single optical depth~\cite{Mahmoud:2018vxx,Kawamura:2021kpx}. For example, some studies fit the spectra of certain sources with two Comptonization components, finding better fits~\cite{Zdziarski:2021nem}. In the soft state, we can expect that the returning radiation of the thermal emission of the disk can be the main source for the production of the reflection spectrum~\cite{Connors:2021nml,Zdziarski:2024jcp}: in this case, the spectrum illuminating the disk would be the combination of thermal spectra with different temperatures and redshifts.

The \texttt{xillverCp} table requires the following input values:
\begin{enumerate}
    \item Electron density ($n_e$) at the disk surface, which we obtain by scaling the gas density in the simulations. We take an average value of the electron density between $\tau = 1$ and $\tau = 10$. This gives a radial density profile along the surface of the disk.
    \item Ionization parameter ($\xi$) at the disk surface, which is calculated for every annulus as
    \be
    \xi_i = \frac{4\pi F_i}{n_{e,i}}.
    \ee
    This provides a radial ionization profile along the disk surface.
    Furthermore, we apply a correction to the calculated ionization profile because \texttt{xillver} spectra are only calculated for a $45\degree$ incident angle \cite{Dauser:2013xv},
    \be
    \xi_{\rm eff} = \xi_i ~\frac{\cos(45\degree)}{\cos(\delta_i)}
    \ee
    where, $\delta_i$ is the average incident angle of the photon on the annulus $i$ calculated using ray tracing (see Eq. (20) in Ref.\cite{Dauser:2013xv}). For brevity, any further mention of ionization is denoted by $\xi$ instead of $\xi_{\rm eff}$.
    \item The emission angle ($\theta_e$), calculated from the ray tracing as
    \be
    \cos(\theta_e) = \frac{k_{\mu} n^{\mu}}{k_{\nu} u^{\nu}}
    \ee
    where, $n^{\mu}$ is the normal vector on the surface of the disk.
    \item Other parameter values are set as follows: photon index $\Gamma = 1.7$, iron abundance $A_{\rm Fe} = 1$, and electron temperature $kT_e = 60~\mathrm{keV}$. We freeze the reflection fraction to $-1$, as we are calculating only the reflection from the disk.
\end{enumerate}
We assume each point on the surface of the disk to be emitting one reflection spectrum. The obtained local spectra are then convolved with the redshift for the distant observer. By taking the emissivity as obtained for the jet corona, we integrate to obtain the relativistic reflection spectrum for the whole disk.

\subsection{Simulating an observation}\label{sec:simulating-observation}

After obtaining the relativistic reflection spectra by convolving ray tracing data with local spectra modeled with \texttt{xillver}, we simulate some observations using the \texttt{fakeit} command in \texttt{XSPEC}~\cite{xspec}. We simulate observations of bright Galactic black holes with NuSTAR~\cite{NuSTAR:2013yza}\footnote{The response, ancillary, and background files used for the NuSTAR simulations were downloaded from \url{https://www.nustar.caltech.edu/page/response-files}.}. 
We assume the energy flux of $\sim 2 \cdot 10^{-8}$~erg~s$^{-1}$~cm$^{-2}$ in the energy band 2-10~keV and observations with exposure time 30~ks (for a single FPM detector).
When simulating the observations, we provide the following model to \texttt{fakeit}:

\vspace{0.2cm}

\noindent {\tt tbabs$\times$(cflux$\times$nthComp + cflux$\times$reflection)} .

\vspace{0.2cm}

\noindent We use \texttt{tbabs}~\cite{Wilms:2000ez} to describe the Galactic absorption and set the hydrogen column density to $N_H = 0.6 \times 10^{22} \mathrm{cm}^{-2}$. \texttt{nthComp}~\cite{Zdziarski:1996wq,Zycki:1999cm} describes the Comptonized component coming from the corona. \texttt{reflection} is the relativistic reflection spectrum obtained from the GRMHD-simulated disk; \texttt{cflux} is used to regulate the relative strength between the Comptonized and reflection components: we impose that the two components have the same power in the energy range 20-40~keV, which essentially mimics a reflection fraction of 1.
In \texttt{nthComp}, we set the photon index $\Gamma=1.7$, electron temperature $kT_e=60~\mathrm{keV}$, and the input disk blackbody temperature $kT_{bb}=0.01~\mathrm{keV}$ in order to be consistent with the values of the same parameters in {\tt reflection}.

\section{Models}\label{sec:models}

\begin{figure*}
    \centering
    \includegraphics[width=0.95\textwidth]{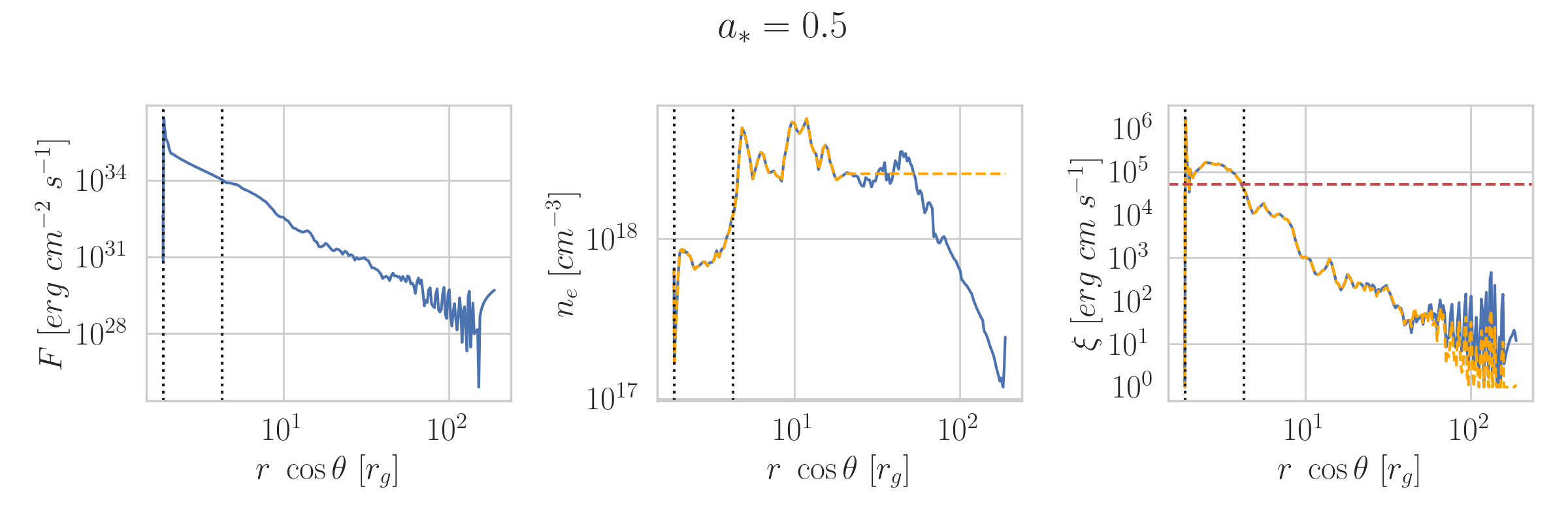}
    \includegraphics[width=0.95\textwidth]{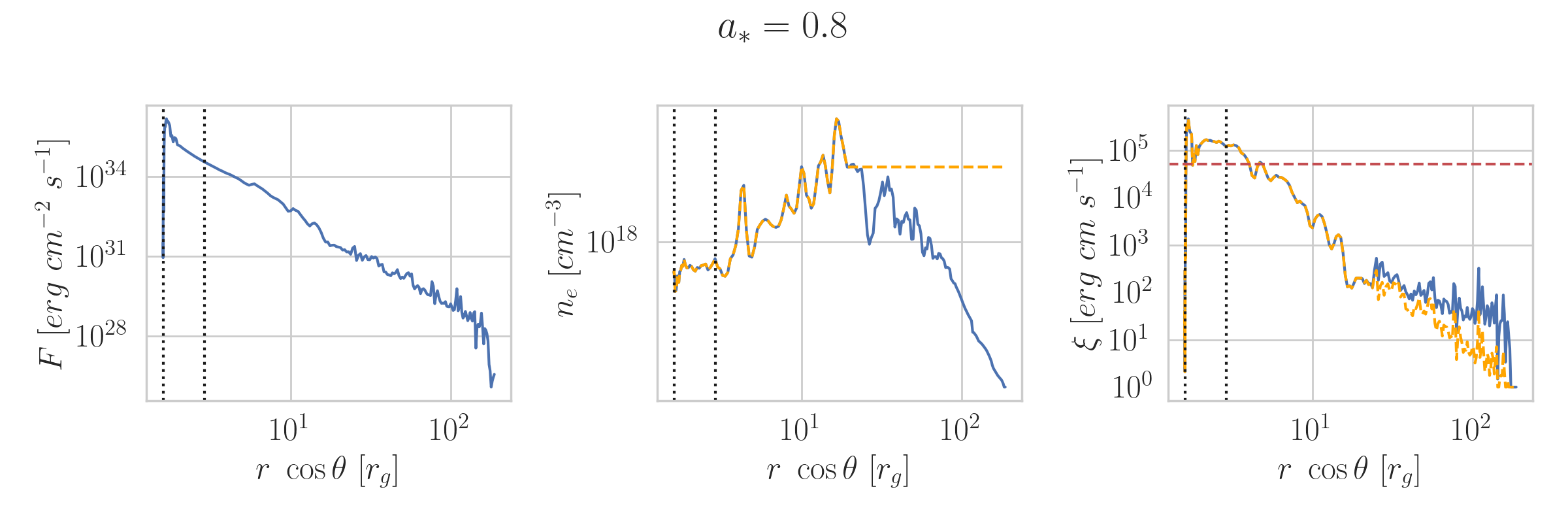}
    \includegraphics[width=0.95\textwidth]{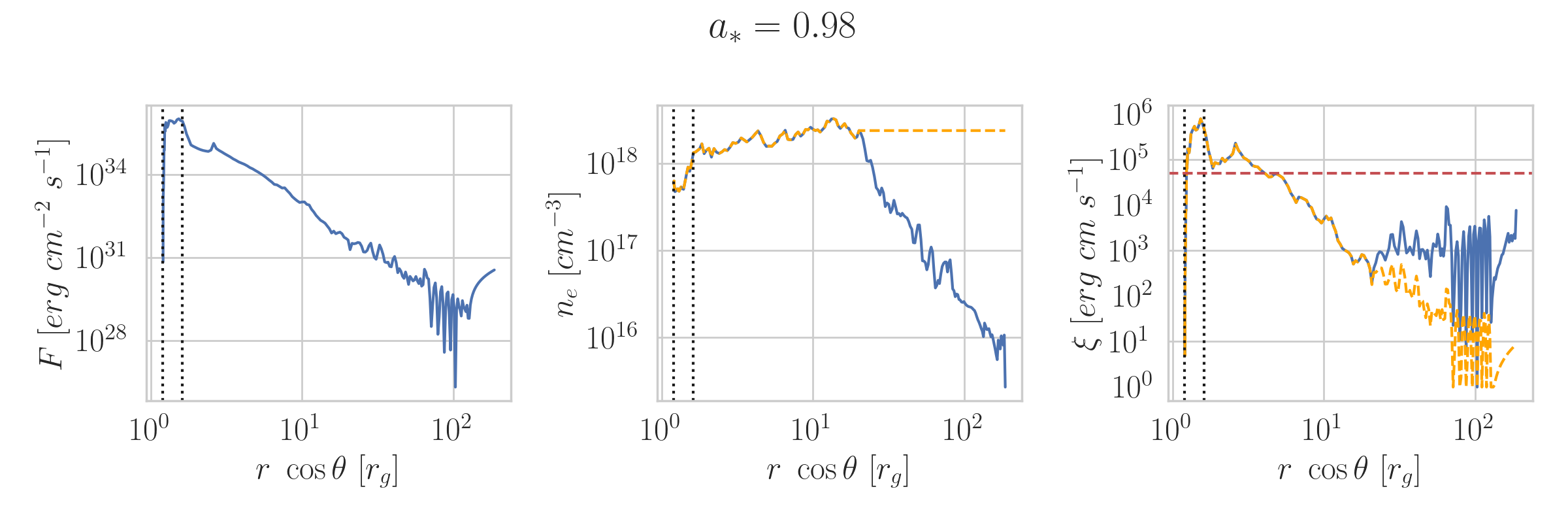}
    \caption{Photon flux from the corona illuminating the disk (left columns), electron density (central columns), and ionization parameter (right columns) obtained from ray tracing onto GRMHD disks for different spins. The black dotted vertical lines mark the radial values of the horizon ($r_{\rm H}$) and the innermost stable circular orbit ($r_{\rm ISCO}$). In the right columns, the red dashed horizontal lines show the highest value of ionization in the \texttt{xillverCp} model viz. $\log\xi = 4.7$. In the central and right columns, the orange dashed curves denote the electron density and ionization parameter profiles when the electron density profile is calculated from the GRMHD simulations for $r < 20~r_g$ and is constant with the value at $r = 20~r_g$ for larger radii (setup M6).}
    \label{fig:emis_ion}
\end{figure*}

From the GRMHD simulations, we obtain the following properties of the accretion disk: $(i)$~the geometry of the disk, $(ii)$~the geometry of the jet, $(iii)$~the four-velocity of the gas, and $(iv)$~the density of the gas from which we calculate the electron density. With the ray tracing code \texttt{blacklamp}, we obtain the emissivity profile produced by the jetlike corona. With the ray tracing code \texttt{blackray}, we obtain the redshift from the surface of the disk to the distant observer and the emission angles at the surface of the disk. With the flux coming from the corona and electron density, we calculate the ionization. We show the coronal flux profile, density profile, and ionization profile in Fig.~\ref{fig:emis_ion}. The quality of data for higher radial values is not very good due to the fact that both GRMHD simulations and ray tracing use logarithmic distributions of radial points to have more points in the inner region of the disk. This does not have any effect on the calculation of the reflection spectra, as most of the reflection radiation is emitted from the inner part of the disk.

After obtaining the spectra for the simulated observations, we fit the data using \texttt{XSPEC}~v12.15.0 with the model \texttt{tbabs$\times$(nthComp + reflection)}.
For \texttt{reflection}, we choose different models within the \texttt{relxill}~v2.5 (v2.5 is same as v2.4 with some emission angle corrections, see Ref.~\cite{Huang:2025fkq}) and \texttt{relxill\_nk}~v1.6 packages. We tabulate the summary of our simulated models and fitting models for \texttt{reflection} in Tab.~\ref{tab:model_cases} for each case. The details for all different cases are as follows:

\begin{itemize}
    \item M0a: This is the simplest model: we simulate a NuSTAR observation with \texttt{relconv$\times$xillverCp} and then fit the simulated data with the same model. This reveals the responses of the instruments.
    
    \item M0b: We calculate the reflection spectra for a Novikov-Thorne disk with our ray tracing code \texttt{blackray}. We choose to set all emission angles equal to the disk's inclination angle. We fit the simulated data with \texttt{relconv$\times$xillverCp} as in the previous case. We use \texttt{relconv$\times$xillverCp} instead of \texttt{relxillCp} specifically to remove any effect of emission angles on the reflection spectra.
    
    \item M1a: We start introducing the details from GRMHD simulations from this model. This model reproduces the case presented in \pone, but here we consider even lower black hole spins. This model introduces the geometry of the disk and the four-velocity of the gas obtained from the GRMHD simulations. The disk is truncated at $r_{\rm ISCO}$. The emissivity profile is described by a power-law with emissivity index $q = 6$. The ionization parameter and the electron density are constant over the whole disk and set, respectively, to $\log\xi = 3.1$ and $\log n_e = 15$ ($\xi$ in units erg~cm~s$^{-1}$ and $n_e$ in units cm$^{-3}$). We fit the data with \texttt{relconv$\times$xillverCp}.
    
    \item M1b: Same as M1a but with $\log n_e = 18$.
    
    \item M1c: Same as M1b but with $q = 3$.

    \item M2a: We calculate the emissivity profile from the jet. The disk is still truncated at $r_{\rm ISCO}$ and ionization and electron density are still constant. To fit the simulated spectra, we use the lamppost model \texttt{relconv\_lp$\times$xillverCp}.

    \item M2b: The simulated model is the same as M2a, but we fit the data with a broken power-law emissivity using \texttt{relconv$\times$xillverCp}.

    \item M2c: The simulated model is the same as M2a, but we fit with twice broken power-law emissivity using \texttt{relconv\_nk$\times$xillverCp}.

    \item M3a: We introduce the effect of emission angle on the simulated spectrum along with jet emissivity, geometry of the disk, and the four-velocity of the gas obtained from GRMHD simulations. The disk is still truncated at $r_{\rm ISCO}$ and ionization and electron density are fixed. We fit the simulated spectra with the \texttt{relxillCp} model with a broken power-law emissivity.

    \item M3b: Same as M3a but we fit the simulated spectra with the lamppost model \texttt{relxilllpCp}.

    \item M4a: We introduce all the calculations from the GRMHD simulations, viz., disk geometry, four-velocity of the gas, jet emissivity, emission angles, plunging region, and ionization and electron density profiles (see Fig.~\ref{fig:emis_ion}) in the simulated spectra. When calculating the simulated spectra, in the case where the ionization $\log\xi > 4.7$, we fix it at 4.7 as that is the maximum value of ionization in the \texttt{xillverCp} table. Note that we do not assume that the disk is truncated at $r_{\rm ISCO}$ any longer: the accretion flow turns out to be optically thick up to the black hole, but there is a {\it plunging region} between the disk and the black hole, where the radial velocity of the gas is high, the density is thus low, and the ionization parameter is high. In the plunging region, the reflection spectrum turns out to be without features because the material is almost fully ionized. We then fit the simulated spectra with \texttt{relxillCp} employing a broken power-law emissivity.

    \item M4b: The simulated model is the same as M4a but we fit using the model \texttt{relxilllpCp} with a fixed ionization and density.

    \item M4c: The simulated model is the same as M4a but we fit with \texttt{relxilllpCp} with a power-law ionization profile and fixed density.

    \item M4d: The simulated model is the same as M4a but we fit with \texttt{relxilllpCp} with a density profile derived from Shakura-Sunyaev $\alpha$-disk and the ionization calculated self-consistently from this density profile.

    \item M5a: Same as M4a, but here we ignore the emission from the annuli with $\log\xi > 4.7$.

    \item M5b: Same as M4b, but we ignore the emission from the annuli with $\log\xi > 4.7$.

    \item M6a: Same as M4a, but we set a constant electron density value for $r > 20~r_g$.

    \item M6b: Same as M4b, but we set a constant electron density value for $r > 20~r_g$.
    
\end{itemize}

\begin{table*}
    \centering
    \caption{Simulated and fitting models for different cases. $\epsilon$ denotes the emissivity profile. For the input $\epsilon$, either we assume a power-law emissivity profile with emissivity index $q$ (i.e., $\epsilon \propto r^{-q}$) or we calculate the emissivity profile from the GRMHD jet. For the fitting $\epsilon$, we can model the emissivity profile with a simple power law (PL), broken power law (BPL), and twice broken power law (TBPL), or with a lamppost emissivity profile. $\xi$ denotes the ionization parameter. In synthetic spectra, we have four cases: $\xi$ is assumed to be constant over the disk (M0-M3), $\xi$ is calculated from the GRMHD disk density profile with the maximum value $\log\xi=4.7$ (M4), $\xi$ is calculated from the GRMHD disk density profile but we ignore the emission when $\log\xi>4.7$ (M5), and $\xi$ is calculated from the GRMHD disk density profile for $r < 20~r_g$ and from the constant electron density value for $r > 20~r_g$. In the fits, we can assume $\xi$ constant over the disk, employ a power law profile (PL), or an $\alpha$ disk profile. $\theta_e$ are the emission angles that are either taken equal to the inclination angle ($i$) of the observer or calculated through ray tracing (RT). The input electron density can be constant over the disk, calculated from the GRMHD simulations, or calculated from the GRMHD simulations for $r < 20~r_g$ and constant for $r > 20~r_g$ (this last case is indicated with G+const. in the table).}
    \begin{tabular}{ccccccccc}
        \hline\hline
        Setup & Simulated Model & Input $\epsilon$ & Input $\log\xi$ & Input $\theta_e$ & Input $\log n_e$ & Fitting Model & Fitting $\epsilon$ & Fitting $\log\xi$\\
        \hline
        M0a & \texttt{relconv$\times$xillverCp} & $q=6$ & 3.1 & $\theta_e=i$ & 18 & \texttt{relconv$\times$xillverCp} & PL & Const. \\
        M0b & \texttt{blackray} & $q=6$ & 3.1 & $\theta_e=i$ & 18 & \texttt{relconv$\times$xillverCp} & PL & Const. \\
        M1a & GRMHD & $q=6$ & 3.1 & $\theta_e=i$ & 15 & \texttt{relconv$\times$xillverCp} & PL & Const. \\
        M1b & GRMHD & $q=6$ & 3.1 & $\theta_e=i$ & 18 & \texttt{relconv$\times$xillverCp} & PL & Const. \\
        M1c & GRMHD & $q=3$ & 3.1 & $\theta_e=i$ & 18 & \texttt{relconv$\times$xillverCp} & PL & Const. \\
        M2a & GRMHD & Jet & 3.1 & $\theta_e=i$ & 18 & \texttt{relconvlp$\times$xillverCp} & Lamppost & Const. \\
        M2b & GRMHD & Jet & 3.1 & $\theta_e=i$ & 18 & \texttt{relconv$\times$xillverCp} & BPL & Const. \\
        M2c & GRMHD & Jet & 3.1 & $\theta_e=i$ & 18 & \texttt{relconv\_nk$\times$xillverCp} & TBPL & Const. \\
        M3a & GRMHD & Jet & 3.1 & RT & 18 & \texttt{relxilllpCp} & Lamppost & Const. \\
        M3b & GRMHD & Jet & 3.1 & RT & 18 & \texttt{relxillCp} & BPL & Const. \\
        M4a & GRMHD & Jet & GRMHD & RT & GRMHD & \texttt{relxillCp} & BPL & Const. \\
        M4b & GRMHD & Jet & GRMHD & RT & GRMHD & \texttt{relxilllpCp} & Lamppost & Const. \\
        M4c & GRMHD & Jet & GRMHD & RT & GRMHD & \texttt{relxilllpCp} & Lamppost & PL \\
        M4d & GRMHD & Jet & GRMHD & RT & GRMHD & \texttt{relxilllpCp} & Lamppost & $\alpha$ disk \\
        M5a & GRMHD & Jet & Ignore $>4.7$ & RT & GRMHD & \texttt{relxillCp} & BPL & Const. \\
        M5b & GRMHD & Jet & Ignore $>4.7$ & RT & GRMHD & \texttt{relxilllpCp} & Lamppost & Const. \\
        M6a & GRMHD & Jet & GRMHD & RT & G$+$const. & \texttt{relxillCp} & BPL & Const. \\
        M6b & GRMHD & Jet & GRMHD & RT & G$+$const. & \texttt{relxilllpCp} & Lamppost & Const. \\
        \hline\hline
    \end{tabular}
    \label{tab:model_cases}
\end{table*}

\begin{figure*}
    \centering
    \includegraphics[width=0.95\textwidth]{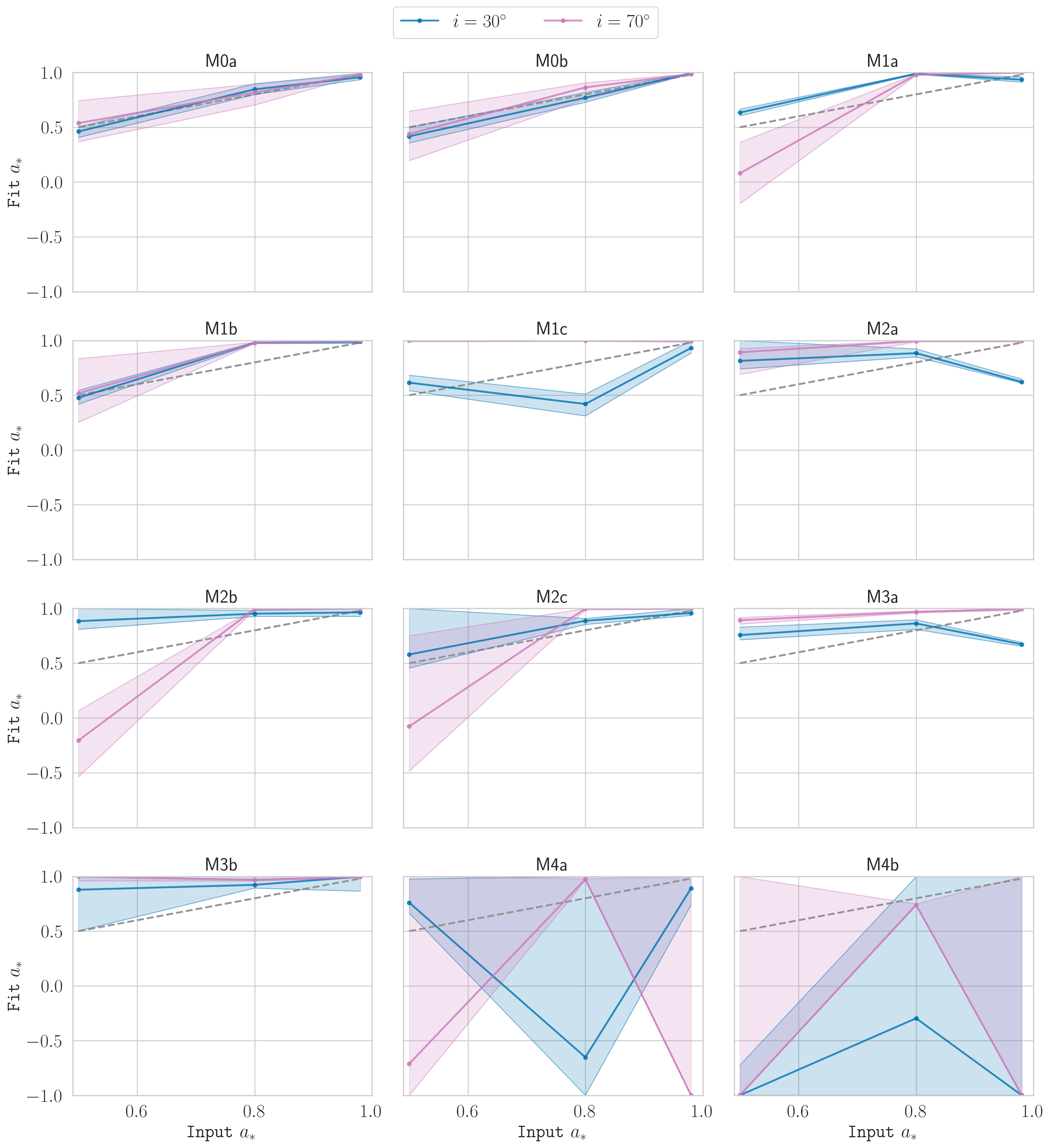}
    \caption{Input spin vs. best-fit spin for cases M0, M1, M2, M3, M4a, and M4b.
    The input spins in our simulations are $a_* = 0.5$, 0.8, and 0.98.
    The best-fit values of the spins are marked by the dots and the shaded regions denote the uncertainties at the 90\% confidence level. The gray dashed line corresponds to the case input spin = best-fit spin.}
    \label{fig:grmhd-models}
\end{figure*}

\begin{figure*}
    \centering
    \includegraphics[width=0.95\textwidth]{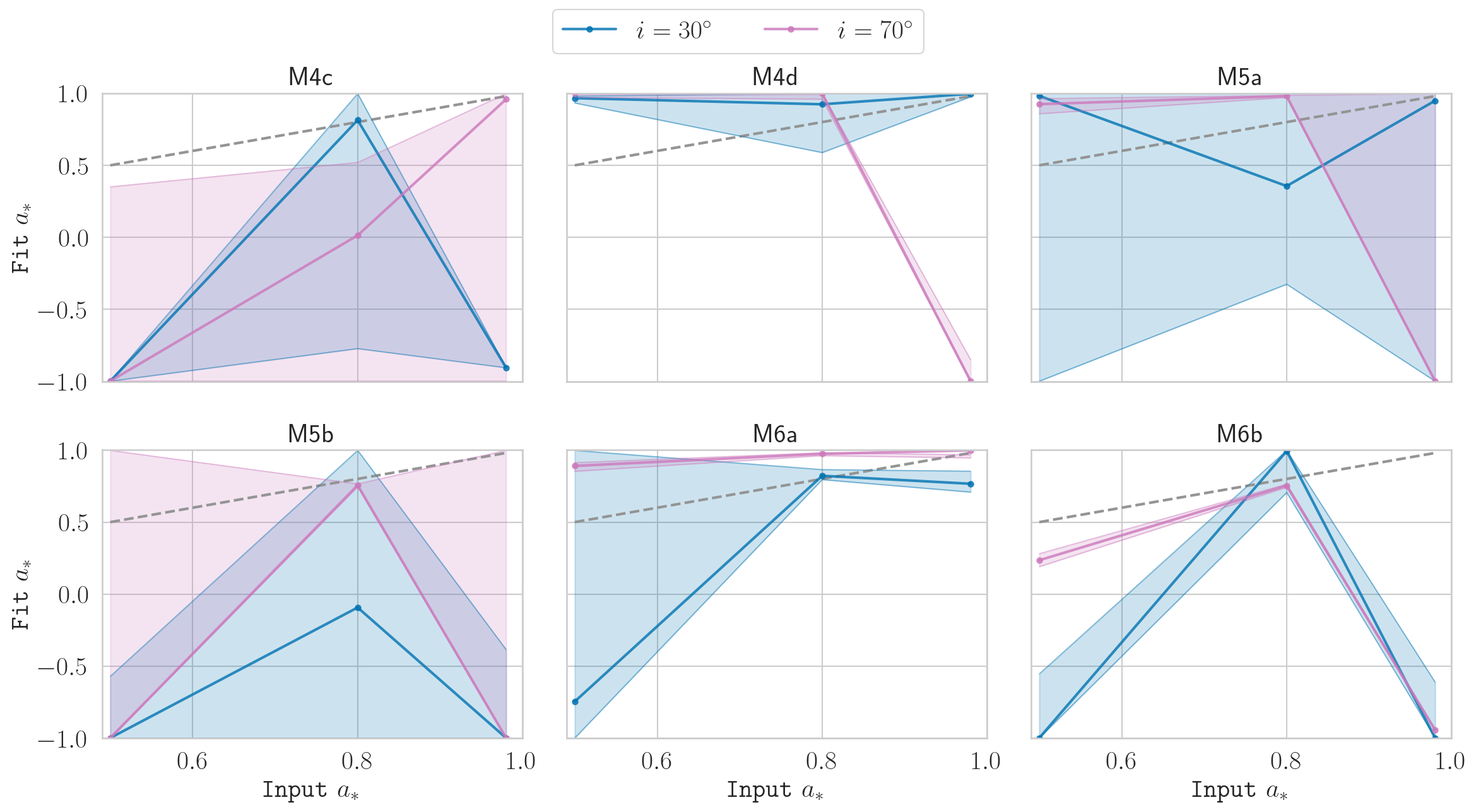}
    \caption{As in Fig.~\ref{fig:grmhd-models} for cases M4c, M4d, M5, and M6.
    The input spins in our simulations are $a_* = 0.5$, 0.8, and 0.98.
    The gray dashed line corresponds to the case input spin = best-fit spin.}
    \label{fig:test-models}
\end{figure*}

\begin{figure*}
    \centering
    \includegraphics[width=0.99\textwidth]{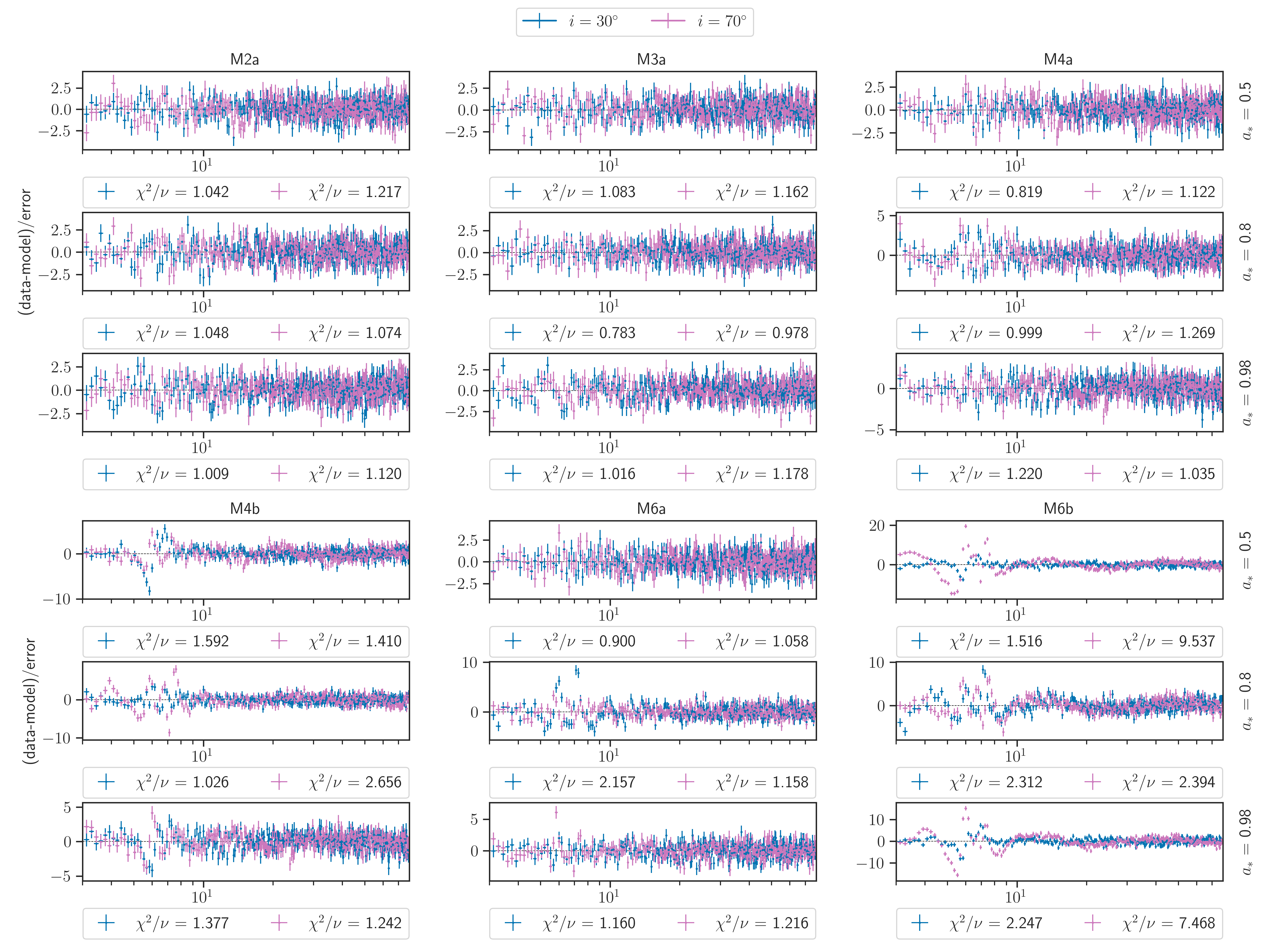}
    \caption{Residuals of the fits of the most representative setups. See the text for more details.}
    \label{fig:delchi}
\end{figure*}

\section{Results}\label{sec:results}

In this work, we tested the state-of-the art of relativistic reflection models against GRMHD simulations of thin accretion disks around black holes. We mainly studied the capability of the reflection models to recover the input spin parameters since spin measurements are one of the important objectives of x-ray reflection spectroscopy. The main idea is to simulate an accreting black hole using GRMHD simulations, which provide a more advanced description of accretion disks compared to analytical disk models, then simulate an observation of $\sim1$~Crab bright Galactic source with an exposure of 30~ks with NuSTAR and then fit the data with our reflection models to see if they recover the input spins.

To test the models, we used several cases (Tab.~\ref{tab:model_cases}) with increasing complexity. Here, we discuss the results of the fit obtained for each case. Fig.~\ref{fig:grmhd-models} and Fig.~\ref{fig:test-models} summarize the results of the spin measurements. The best-fit values of the most representative setups are reported in Supplemental Material\cite{supplementalMaterial}.

\subsection{Setup M0}

In M0a and M0b, we test our method (Fig.~\ref{fig:grmhd-models}). For M0a, we simulate the spectrum using the \texttt{relconv$\times$xillverCp} model and then fit the simulated observation with the same model. This reveals the response of the instruments on board NuSTAR. We see that the input spins are recovered as expected and that the uncertainties are higher for lower spins. In M0b, we test our code \texttt{blackray}. The code should produce the same spectrum as \texttt{relconv$\times$xillverCp}, and we find that the fitting produces very similar results as M0a. It has to be remarked that we use \texttt{relconv$\times$xillverCp} and not \texttt{relxillCp} in our simplest cases because in \texttt{relxillCp} there are additional calculations related to emission angles that are taken into account (see Ref.~\cite{Huang:2025fkq}), so choosing a convolution model removes that added complexity.

\subsection{Setup M1}

In the three cases for M1a-M1c (Fig.~\ref{fig:grmhd-models}), we use GRMHD data of the geometry and four-velocity of gas from the GRMHD simulations and use fixed values of emissivity index, ionization and electron density. M1a basically replicates the result of \pone, but here we include even lower spin values. M1b is the same as M1a but with a higher electron density of $\log n_e = 18$ compared $\log n_e = 15$ for M1a. In both cases, we see that the spin is almost recovered for $a_*=0.5$ and $a_*=0.98$ with tighter constraints for higher spins. The spin is overestimated for the model with $a_*=0.8$. In model M1c, we reduce the emissivity index to $q=3$ (it was $q=6$ in the previous cases\footnote{Power-law emissivity is defined as $\epsilon \propto r^{-q}$}). This reveals that the spin is always overestimated for lower inclinations. The higher inclination the spin is better estimated, with $a_*=0.8$ model being slightly underestimated.

\subsection{Setup M2}

In M2a-M2c (Fig.~\ref{fig:grmhd-models}), we study the impact of the emissivity profile of the jet. The lamppost geometry is usually thought to describe a system in which the corona is the base of the jet. In our case, we take the base of the jet from the GRMHD simulations as an extended corona and calculate the reflection spectra produced by such an emissivity. We fit the spectra with a lamppost model in M2a, with a broken power-law emissivity profile in M2b, and a twice broken power-law emissivity profile in M2c. 
We find that M2b and M2c can recover the correct spin for $a_*=0.98$, regardless of the inclination angle of the disk. M2b fails to recover the correct spins for lower values. M2c recovers almost the correct spin for all cases. In the case of M2a (lamppost), we can never recover the correct input spin and the model cannot fit the data well: we find large residuals at the iron line (see Fig.~\ref{fig:delchi}). We note that the spin is recovered better when using a twice broken power-law model for the extended jetlike corona. However, we do not have any further results with twice broken power-law emissivity because the \texttt{relxill} models do not have it and \texttt{relxill\_nk} currently lacks a high density model with a \texttt{nthComp} incidence spectrum.

\subsection{Setup M3}

In M3a and M3b (Fig.~\ref{fig:grmhd-models}), we fit with models that take into account the emission angle at the disk, as in the previous models the emission angle was taken to be the same as the inclination angle. There are no major differences between the fits of models M3a from M2a. M3b performs better than M2b for the case $a_*=0.5$ and $i=70\degree$. The lamppost model cannot recover the correct spins, while the model with a broken power-law emissivity profile can at least recover the correct spins in the case of fast-rotating black holes. We conclude that the effect of emission angle is small when considering current instruments like NuSTAR. This result is in line with our recent study in Ref.~\cite{Huang:2025fkq}.

\subsection{Setup M4}

In cases M4a-M4d, we use the most complex models derived from the GRMHD simulations: the electron density and ionization profiles are inferred from the GRMHD and ray tracing simulations (as shown in Fig.~\ref{fig:emis_ion}) along with the disk and corona geometries used in the previous cases (see Fig.~\ref{fig:grmhd-models} for M4a and M4b and, Fig.~\ref{fig:test-models} for M4c and M4d). This provides us with the most sophisticated model in our study. While fitting, in M4a we use \texttt{relxillCp} with a broken power-law emissivity profile and in M4b we use \texttt{relxilllpCp} with the lamppost profile. We see both models fail to constrain the spin. M4a performs slightly better in fitting, as we see more residuals in M4b around the iron line (see Fig.~\ref{fig:delchi}). \texttt{relxilllpCp} has options to choose ionization and density profiles using the parameter \texttt{iongrad\_type}. In M4c, we use a power-law ionization profile: this model is not able to fit and constrain the spin (Fig.~\ref{fig:test-models}). In M4d, we use the $\alpha$-disk electron density profile, which self-consistently calculates the ionization profile. M4d can constrain the spins, but it cannot recover the correct values. We note that:  $(i)$ the density profile in our simulations is closer to a Novikov-Thorne disk profile, and $(ii)$ as in the previous cases, the lamppost model cannot fit the jetlike corona. However, as shown below with case M6, the M4 cases fail to measure the spins because of the small size of the disk in our GRMHD simulations.

\subsection{Setup M5}

Since cases M4a-M4d fail to constrain the spins correctly, with M5 and M6 we try to figure out the reason of such discrepancies (Fig.~\ref{fig:test-models}). In M5, we ignored the highly ionized emission from the inner regions of the disk, so we ignore the emission when $\log\xi > 4.7$. In M5a, we fit the simulated data with a broken power-law emissivity profile. In M5b, we use the lamppost profile. Both models, like their corresponding M4a and M4b, failed to constrain the spins and produced similar results (Fig.~\ref{fig:test-models}). This result shows that the emission from the plunging region is not the reason preventing the recovery of the correct input spins.

\subsection{Setup M6}

In M6a and M6b (Fig.~\ref{fig:test-models}), we consider the fact our simulated disks are small for computational reasons and their densities drop unnaturally fast for $r > 20~r_g$. We thus use the electron density from the GRMHD simulations for $r < 20~r_g$ and the electron density at $r = 20~r_g$ for $r > 20~r_g$. This decreases the ionization parameter value at $r > 20~r_g$ and its radial profile appears to be consistent with the region $r < 20~r_g$ (see the orange dashed profiles in the right panels in Fig.~\ref{fig:emis_ion}). In M6a with a broken power-law emissivity profile, we recover the correct input spins for $a_* = 0.8$ with $i = 30 \degree$ and for $a_* = 0.98$ with $i = 70 \degree$, the spin is unconstrained for $a_* = 0.5$ with $i = 30 \degree$, and we do not recover the correct spins in the other cases. With the lamppost emissivity profile in M6b, we cannot fit well the data and we see large residuals in Fig.~\ref{fig:delchi}.

\section{Conclusions}\label{sec:conclusion}

In this work, we tested the capability of relativistic reflection models to measure the spins of black holes from simulated data produced by GRMHD and ray tracing techniques. Our main findings are as follows: $(i)$ we recover the correct black hole spin for $a_* = 0.98$ and $i = 70 \degree$ with a model employing a broken power-law emissivity profile; $(ii)$ we find some discrepancy between the input spin and the spin measurement in the case of $a_* = 0.98$ and $i = 30 \degree$ with a model employing a broken power-law emissivity profile; we note that $(i)$ and $(ii)$ are consistent with the conclusions of \pone, even if our model here is more advanced (we calculate the emissivity profile, electron density profile, and ionization profile from the GRMHD simulations); $(iii)$ we find large biased spin measurements if the black holes are not fast rotating (input spins $a_* = 0.5$ and 0.8) with a model employing a broken power-law emissivity profile; $(iv)$ the models with lamppost emissivity profile are unsuitable to fit our simulated data in which the corona is described by the jet inferred from the GRMHD simulations.

We focused only on the spin constraints. However, there can be degeneracies with other parameters which could result in incorrect measurements of spins. Studying the correlations among different parameters is beyond the scope of this work. In Supplemental Material\cite{supplementalMaterial}, we provide the tables with the best fits for our different setups. We find that an incorrect spin measurement may be related to the estimate of the parameters related to the emissivity profile (i.e., height of the corona or power-law indices) and to some degree the inclination angles. We also find that in some cases, when the estimate of the electron temperature $kT_e$ is lower than in the simulations, the spin is underestimated and often negative. In all our fits, we freeze the value of the hydrogen column density $N_H$ because usually it cannot be constrained from the fit. If we simulate simultaneous observations NICER+NuSTAR, we can constrain $N_H$ and we obtain a better estimate of $kT_e$ (see Appendix and Supplemental Material\cite{supplementalMaterial}).

Our simulations predict that the plunging region is optically thick and highly ionized, which is consistent with the result found in Ref.~\cite{Reynolds:2007rx}. In such a case, the x-ray photons from the corona can produce reflection radiation from the plunging region, but only through Compton scattering and there are no emission features in its spectrum. Such a radiation can thus contribute to the reflection spectrum, but cannot appreciably affect the analysis of the reflection features and, in turn, the measurement of the black hole spin.

Today there is a debate on the strong tension between the spin measurements of stellar-mass black holes from x-ray reflection spectroscopy (where most black holes are found to be fast rotating, say $a_* \gtrsim 0.7$~\cite{Draghis:2023vzj}) and those from gravitational wave observations (where black holes are found to be slow-rotating, say $a_* \lesssim 0.4$); see, for example, Refs.~\cite{Zdziarski:2025ozs, Fishbach:2021xqi}. This discrepancy is confirmed by the latest gravitational wave transient catalogs, which include a larger number of gravitational wave events and black hole spin measurements~\cite{GWTC3,GWTC4}. It may be caused either by the fact that these black holes belong to two different populations or at least one of the two methods does not provide accurate spin measurements.
To further complicate the picture, some authors find that truncated disks are common in the hard states~\cite{2015ApJ...814...50D,2016MNRAS.458.2199B,2017MNRAS.472.4220B,2021ApJ...909L...9Z}, which is not consistent with the high-spin measurements often inferred in the hard state, where reflection features are usually stronger. Some authors have also shown that fitting black hole x-ray data with more complex spectra (e.g., by adding a second Comptonized component) can lead to finding lower spins or truncated disks (see, e.g., Ref.~\cite{2024ApJ...967L...9Z}).

The majority of the known black holes in x-ray binaries are in low-mass x-ray binaries. We only know a few black holes in high-mass x-ray binaries (and only Cygnus~X-1, M33~X-7, LMC~X-1, LMC~X-3, and IC~10~X-1 have spin measurements reported in the literature). In low-mass x-ray binaries, the companion stars have a mass $< M_\odot$, so they cannot evolve into black holes and these systems are not the progenitors of gravitational wave events. The case of high-mass x-ray binaries is more complicated~\cite{Zdziarski:2025ozs}. For example, Cygnus~X-1 is thought to evolve into a binary black hole, but the system may merge after a very long time~\cite{Ramachandran:2025wnd}. M33~X-7 is not supposed to become a binary black hole: the black hole and the companion star will probably merge during a common-envelope stage~\cite{Ramachandran:2022xrh}.

Current models for the gravitational collapse of heavy stars predict that the formation of fast-rotating black holes from stellar collapses are unlikely~\cite{Woosley:2006fn,Yoon:2006fr}, which would be consistent with the results from gravitational wave observations. While it is often thought that black holes in x-ray binaries cannot significantly change the value of their spin~\cite{King:1999aq,Valsecchi:2010cw,Wong:2011eg}, some authors have proposed mechanisms to spin-up black holes, either those in high-mass x-ray binaries~\cite{Qin:2018sxk} or those in low-mass x-ray binaries~\cite{Podsiadlowski:2002ww,Fragos:2014cva}, to explain the high spin values reported in the literature.

In Ref.~\cite{Zdziarski:2025ozs}, the authors point out that the accretion disks predicted by the Novikov-Thorne model are unstable and the latter should not be used for spin measurements. However, reflection models only assume that the accretion disk is perpendicular to the black hole spin axis and the motion of the material in the disk is Keplerian, without employing the actual structure of the Novikov-Thorne disks.

If we consider the most advanced simulations (M3a, M3b, M6a),\footnote{We ignore the setups M4 and M5, because they are strongly affected by the small size of the accretion disk in our simulations, and M6b, because the quality of the fits is poor.} we see that the analysis of the reflection spectrum tends to infer high spin values, especially in the case of a high inclination angle. We can normally infer the correct spin parameter in the case of fast-rotating black holes, but we may also infer high spin values for slow-rotating black holes. Since in the simplest simulations (setup M0) we always recover the correct black hole spins, it is possible that by improving the accretion disk structure and description of the emissivity profile in current reflection models we can also measure the spins of slow-rotating black holes. It is worth noting that similar high estimations of spins also occur with the continuum fitting method for disk thermal spectra obtained from GRMHD simulations, see Refs.~\cite{Wielgus:2022qij,Lancova:2022bha}.

A natural question is how we can distinguish a true fast rotating black hole from a fake fast rotating black hole if we infer a high value of the black hole spin with x-ray reflection spectroscopy. While we have not found any robust quantitative criterion within our simulations, our results support the conclusion already pointed out in Ref.~\cite{Dauser:2013xv}. Very broadened iron lines are only possible in the case of fast rotating black holes when their corona is compact and close to the black hole. If the black hole is fast rotating but the corona is extended or not very close to the black hole {\it or} if the corona is compact and close to the black hole but the black hole is slow rotating, the iron line is not very broadened and these two cases cannot be easily distinguished. If the black hole is slow rotating and the corona is not compact and/or not close to the black hole, relativistic effects on the reflection spectrum may be even too weak to allow us to measure the black hole spin. The conclusion is thus that reliable spin measurements would be possible only for fast-rotating black holes in which the corona is compact and close to the black hole. In all other cases, spin measurements using x-ray reflection spectroscopy are more difficult and the final results should be taken with caution. \\

\begin{acknowledgments}
We thank Christopher~J.~White for an earlier collaboration on this work.
This work was supported by the National Natural Science Foundation of China (NSFC), Grants No.~12250610185 and No. 12261131497. S.S. is supported by the Shanghai Super Postdoctoral Fellowship. Y.M. is supported by the National Key R\&D Program of China (Grant No.~2023YFE0101200), the National Natural Science Foundation of China (Grant No.~12273022), and the Shanghai Municipality orientation program of Basic Research for International Scientists (Grant No.~22JC1410600). The computations in this research were performed using the CFFF platform of Fudan University.
\end{acknowledgments}

\appendix

\section{NICER + NuSTAR fits}

\begin{figure*}
    \centering
    \includegraphics[width=0.95\textwidth]{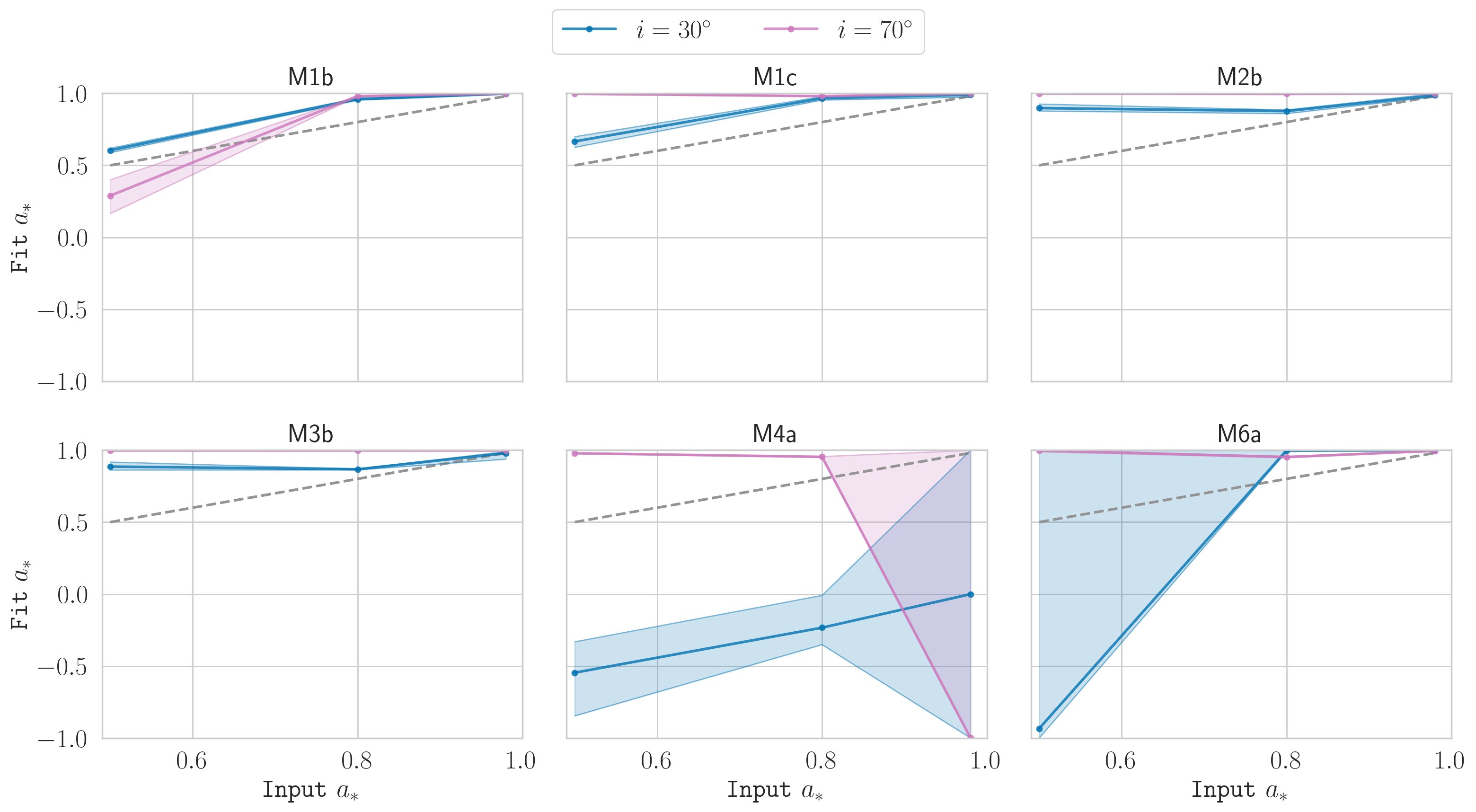}
    \caption{As in Fig.~\ref{fig:grmhd-models} for simultaneous observations NICER+NuSTAR. The gray dashed line corresponds to the case input spin = best-fit spin.    } 
    \label{fig:nicer-nustar}
\end{figure*}

In Sec.~\ref{sec:results}, we assumed observations with NuSTAR. Here we simulate simultaneous observations with NICER+NuSTAR. For NICER~\cite{2016SPIE.9905E..1HG}, we assume an exposure time of 5~ks\footnote{The response, ancillary and background files for NICER were downloaded from \url{https://heasarc.gsfc.nasa.gov/docs/nicer/proposals/nicer_tools.html}.}. For NuSTAR, we assume an exposure time of 30~ks as before. With only NuSTAR, we were unable to constrain the hydrogen column density $N_H$ in \texttt{tbabs} for some cases. With NICER, we can do it. In Fig.~\ref{fig:nicer-nustar}, we show the results for some of the most representative cases. As in the case of the observations with only NuSTAR, the lamppost model does not perform well and therefore we do not present its results here. In all the models, we do not see any particular difference in these new simulations from NICER+NuSTAR concerning spin measurements, and the only difference is that the uncertainties are usually smaller.

\section{Iron lines}

\begin{figure*}
    \centering
    \includegraphics[width=0.98\textwidth]{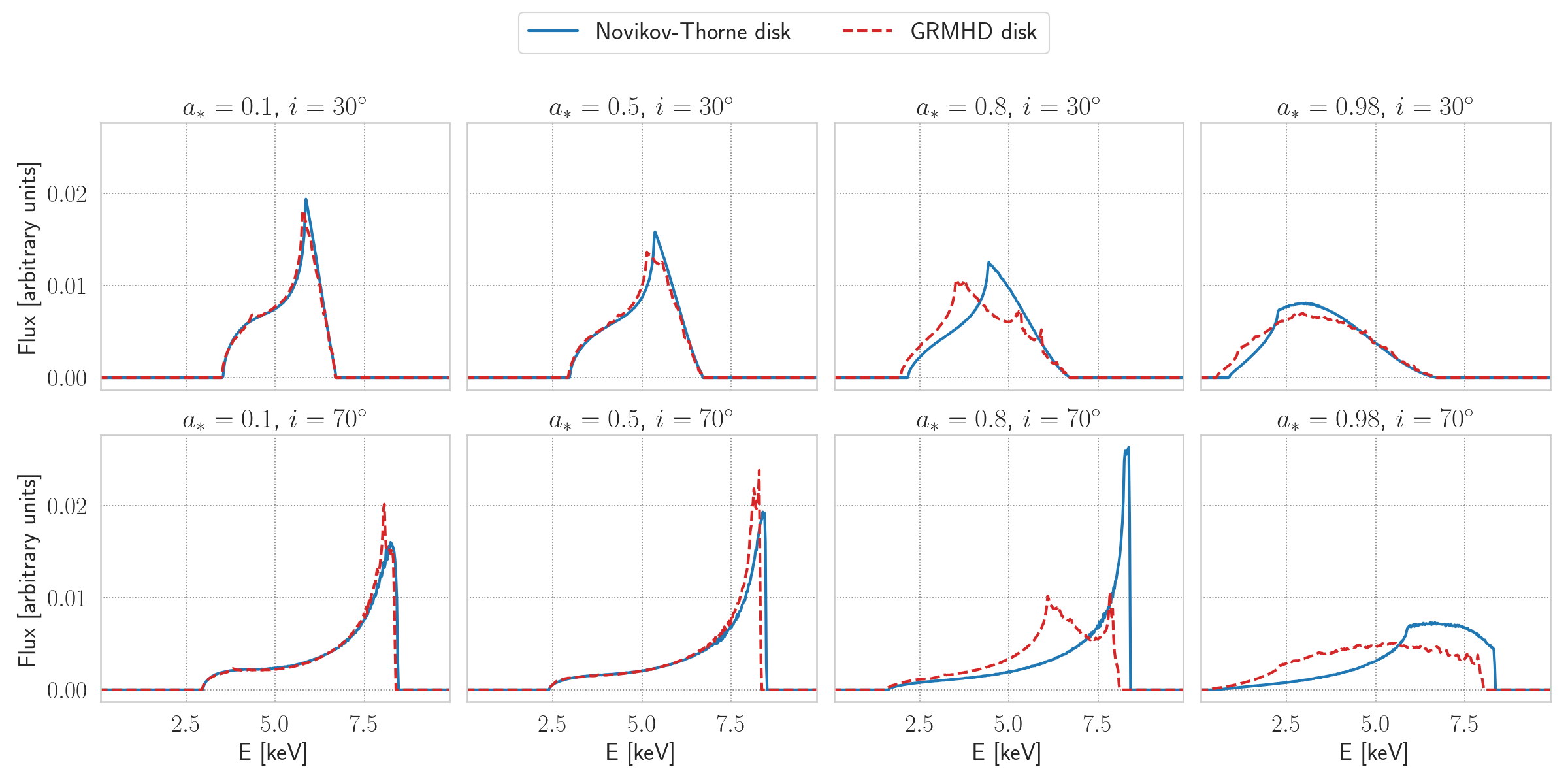}
    \caption{Iron line profiles obtained from the Novikov-Thorne disk model (solid blue curves) and GRMHD disk model (dashed red curves). The emissivity index in both cases is fixed $q=6$. The top panels are for low inclination ($30\degree$) and bottom panels are for high inclination ($70\degree$). The first (left) panels are for spin $a_*=0.1$, second for $a_*=0.5$, third panels for $a_*=0.8$ and fourth (right) panels for $a_*=0.98$. Note that in the case of GRMHD models the disk is truncated at the ISCO.}
    \label{fig:ironlines}
\end{figure*}

In Fig.~\ref{fig:ironlines}, we show the comparison of the iron lines obtained from the Novikov-Thorne disk and GRMHD-simulated disk for fixed emissivity of $q=6$ using our code \texttt{blackray}.

\section{Case with spin $a_* = 0.1$}

\begin{figure*}
    \centering
    \includegraphics[width=0.95\textwidth]{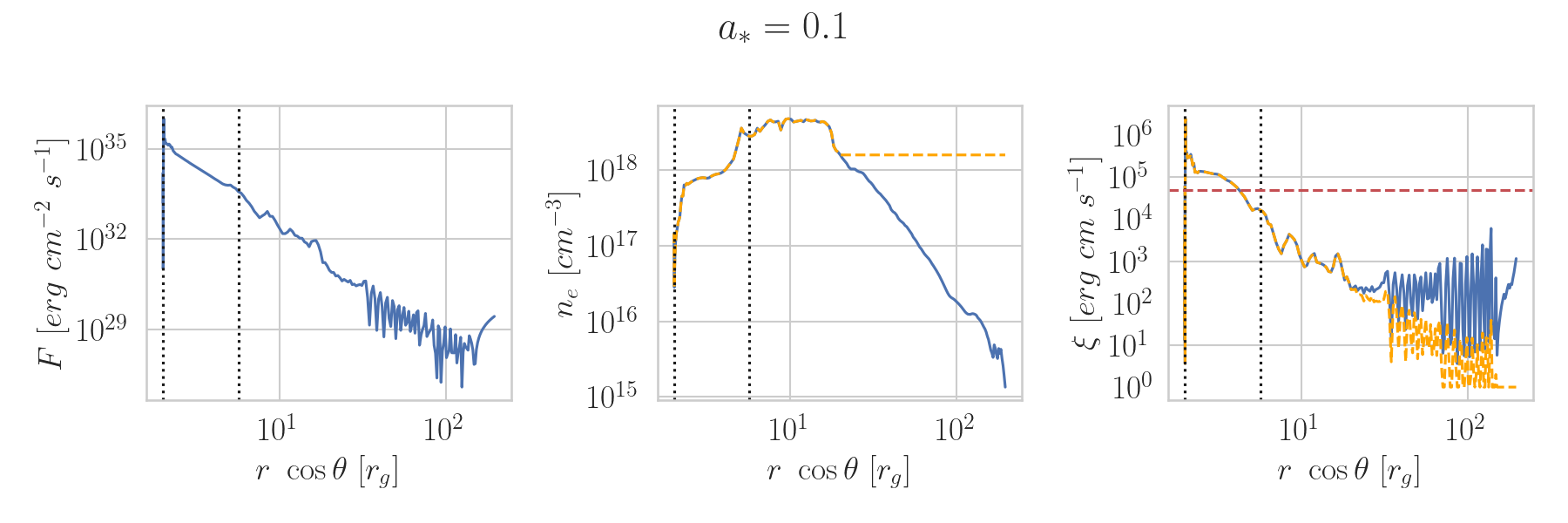}
    \caption{Same as Fig.~\ref{fig:emis_ion} but for spin $a_* = 0.1$.}
    \label{fig:em0p1}
\end{figure*}

\begin{figure}
    \centering
    \includegraphics[width=0.95\linewidth]{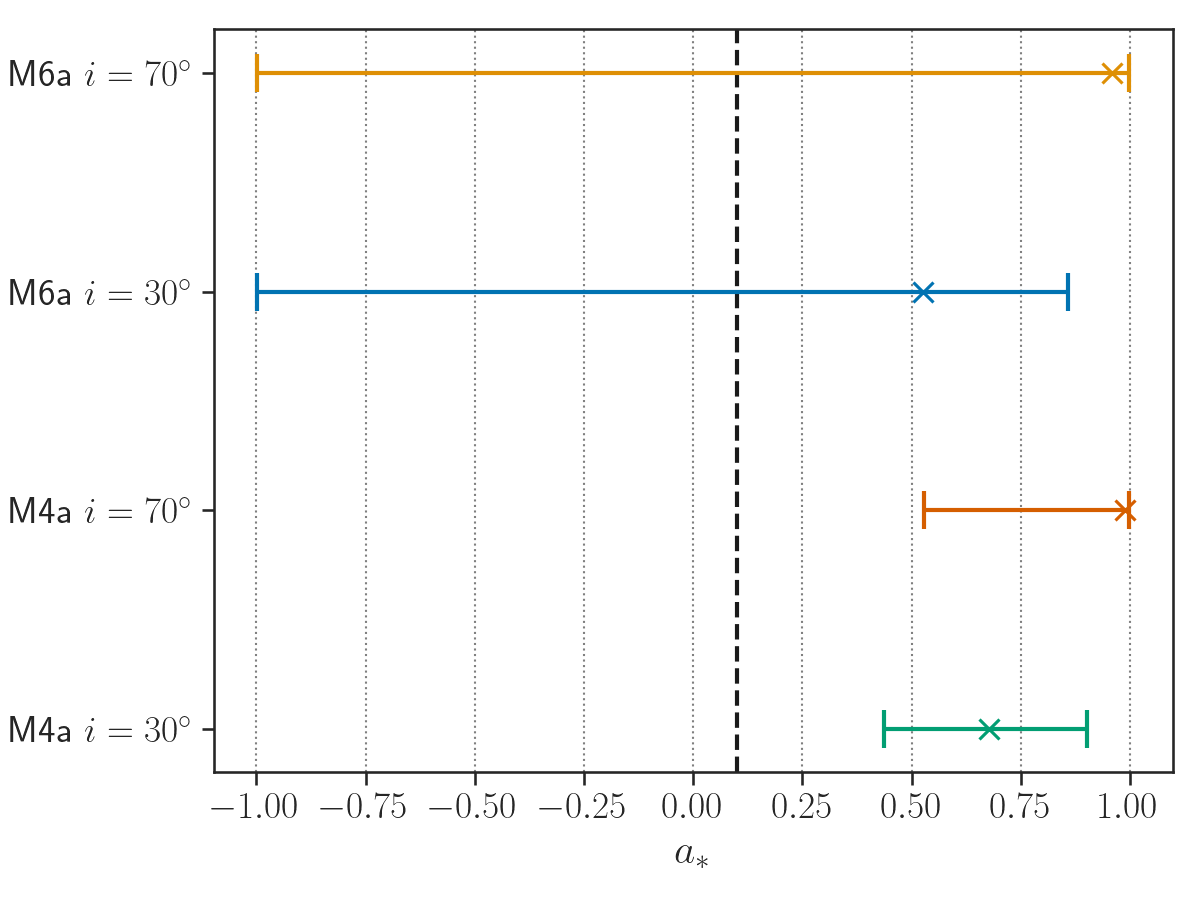}
    \caption{Setups M4a and M6a results with spin $a_* = 0.1$. The bars denote 90\% confidence interval and the cross is the best-fit value.}
    \label{fig:a0p1}
\end{figure}

The results of gravitational wave measurements predict a median value of the BH spin around $a_* = 0.1$. Here we run our analysis for the setups M4a and M6a with input spin $0.1$.
Simulation and ray tracing results are shown in Fig.~\ref{fig:em0p1}.
The results are shown in Fig.~\ref{fig:a0p1} and are similar to other cases, i.e., either high spin is recovered (M4a) or the spin is unconstrained (M6a).

\bibliography{references}

\end{document}